%% file: main.tex
\documentclass{article}

\input{packages}

\input{theorems}
\newcommand{\cum}{\mathrm{cum}}
\newcommand{\RA}{\mathrm{RA}}

\newcommand{\Ret}{\mathrm{Ret}}
\newcommand{\nom}{\mathrm{nom}}

\newcommand{\aitch}{\mathsf{h}}

\newcommand{\YTM}{\mathsf{YTM}}
\newcommand{\SI}{\mathsf{SI}}
\renewcommand{\CPI}{\mathsf{CPI}}
\newcommand{\AWE}{\mathsf{AWE}}
\newcommand{\LTIE}{\mathsf{LTIE}}
\newcommand{\DCPI}{q}
\newcommand{\DCPIt}{q_t}

\DeclareMathOperator{\logsumexp}{logsumexp}
\DeclareMathOperator{\im}{Im}

\title{A comparison of the effectiveness of alternative DC and CDC designs in a UK market}
\author{\small{John Armstrong\textsuperscript{a}, James Dalby\textsuperscript{b}, Rohan Hobbs\textsuperscript{c}}\\
\small{\textsuperscript{a,b,c}Department of Mathematics, King's College London, UK}\\
\small{\textsuperscript{a,*}john.armstrong@kcl.ac.uk, \textsuperscript{b}james.dalby@kcl.ac.uk, \textsuperscript{c}rohan.hobbs@kcl.ac.uk} }

\begin{document}

\maketitle

\begin{abstract}
    We use three stochastic models to evaluate the effectiveness of a number of possible pension designs which have been proposed for use in the UK.
    We consider individual DC schemes followed by full annuitisation and a flex-and-fix strategy which combines drawdown with gradual annuitisation. We compare these approaches with collective designs including: a flat-accrual shared-indexation CDC scheme that is similar to the Royal Mail Collective Pension Plan; a dynamic-accrual shared-indexation CDC scheme modelled on the approach considered in the DWP consultation on multi-employer CDC; and an alternative collective design based on a tontine structure. In our comparisons, we tune
    each strategy to give optimal performance given the stochastic model
    and a choice of representative risk preferences.

    We find the collective design based on a tontine structure consistently achieves the best performance in terms of member utility. We discuss the importance of leverage in the optimal investment strategies.
\end{abstract}

\section{Introduction}

The goal of this paper is to provide a risk-adjusted numerical comparison of
the effectiveness of a number of different possible pension scheme designs
that have been proposed for the UK. Our primary motivation is to assess the potential benefits of collective defined contribution (CDC) designs compared
to individual defined contribution (DC) arrangements.

The term CDC is somewhat ill-defined as quite varied designs have been proposed internationally, see for example \cite{haan, bonenkamp,cui,gollier}. We use the term broadly to mean any design where there is no employer covenant to guarantee payments in retirement, but where members seek to provide each other with some form of collective insurance.

Under current UK legislation, all CDC schemes must follow a similar design which we call {\em shared-indexation}. The first, and currently only, UK CDC scheme was the Royal Mail Collective Pension Plan, launched in 2024 \cite{wilkinson}. It follows a specific form of shared-indexation design we call shared-indexation with {\em flat-accrual}.

However, if one understands ``CDC'' more broadly than UK legislation, many other collective designs are possible. 
We will describe a possible
design for a collective scheme that allows for individual choice and  flexible investment strategies which we call {\em collective drawdown}. This design simply combines a tontine structure for pooling longevity risk with an optimal investment strategy.
It is collective in the sense that longevity risk is pooled and, more subtly, we will see that a collective agreement can enable more cost-effective leveraged investment strategies.
A key feature of the design is that we use a tontine structure
which allows longevity risk to be shared effectively among members who wish to pursue different investment strategies.
We claim that it is possible to prove that, subject to some caveats, most importantly the assumption that the market is complete, the collective drawdown design is optimal. The complete proof is beyond the scope of this paper, but will be provided in a forthcoming manuscript.
However, in this paper we simply seek to quantify the size of this potential benefit.  

We will assess the benefits of both shared-indexation and collective drawdown approaches. We will compare them with the annuity-based approaches of full annuitisation at retirement and {\em flex-and-fix} where one gradually annuitises wealth from retirement while maintaining some risky investment.\footnote{The similar term {\em flex-then-fix} is used by practitioners to refer to a strategies where annuitisation is deferred till a number of year after retirement. The term {\em fix-then-flex} is used to refer strategies with partial annuitisation at retirement. Our term {\em flex-and-fix} includes both of these possible strategies as well as more complex strategies where annuitisation is spread over time.}

It is clear a-priori that collective drawdown design will outperform any DC approaches as the design benefits from more flexibility and full longevity pooling. It is not obvious which will perform better, shared-indexation designs or a flex-and-fix approach. The former is able to combine longevity pooling with risky investment, the latter allows more scope for optimization but only allows longevity pooling through riskless annuities. Our results show that flex-and-fix can be competitive with shared-indexation designs, despite the lack of longevity pooling.

\medskip

Comparisons between shared-indexation designs and traditional DC alternatives have been performed in the past, with varying degrees of modelling sophistication
\cite{aon, wtw, popat, owadally, armstrong_dalby_donnelly}. However, these comparisons are made on the basis
that the schemes follow pre-specified  investment strategies.

By contrast, in this paper we will select the DC designs by solving appropriate optimal investment problems.
This is an important consideration because a weakness of the shared-indexation design is that it provides
little flexibility for optimization. This is for two reasons: firstly, the design is quite tightly specified by regulations; secondly, the shared-indexation design provides no individual choice for members and so cannot be optimized for each member. This limitation of shared-indexation should be incorporated in any comparison of shared-indexation schemes with DC schemes. We are also not aware of any previous comparison of the CDC designs we consider with a {\em flex-and-fix} approach
to DC decumulation.

\medskip

It is clear what it means to optimize a strategy, namely to maximize a single investor's target utility. We identify the optimal flex-and-fix strategy, full annuitisation strategy and collective drawdown strategy by solving the appropriate utility maximization problems using machine-learning techniques.

It is less clear what it means to optimize a shared-indexation scheme. In such a scheme, all the fund's assets are pooled, and so all members will experience the results of any investment decisions. As there are many investors in different generations to consider,  the choice of objective function is non-obvious. We have approached this by optimizing the outcome for the worst-case between the outcome for the first generation to be in the scheme for their entire career, and a generation which joins shortly before the scheme is closed to new entrants. This is intended to capture the worst-case among the generations who experience the ``steady-state'' operation of the scheme.

There are very few parameters that can be adjusted to optimize a shared-indexation design. The key parameter
is the asset mix held in the steady state of the scheme and this can be varied in all the shared-indexation designs. The only other parameter that has a significant effect on performance is called the ``target level for the indexation'' but for some designs the asset mix determines this level. Thus, in the steady state of operation, the shared-indexation designs depend only on a small number of parameters, so we can identify optimal designs by a brute force search.

\medskip

The results of the optimization process inevitably depend upon the choice of model. Broadly speaking,
the richer the model one uses the greater the scope will be
for optimization. There are pros and cons to any model, and so we have repeated our analysis
with three different economic models to provide some insight into how results depend upon this modelling choice. The models used are the same as those considered in \cite{armstrong_dalby_donnelly}.

The first model we consider is a Black--Scholes model. This has the advantage of being the simplest to understand and it provides the most straightforward comparisons. The second model is a stochastic interest-rate (SIR) model but where wage-inflation is assumed constant in real terms. This model has been chosen to add the important feature of fluctuating interest rates while otherwise maintaining parsimony of the Black--Scholes model.
The third model is a rich model containing many risk factors taken from \cite{armstrongMaffraPennanen}. Using this model ensures that a full range of risks are modelled including features such as inflation, wage inflation and economic cycles.

To minimize the potential for over-fitting in the more complex models, we have restricted all funds to invest entirely in equities and
index-linked bonds. This creates a more level playing-field between the different pension schemes as well as helping to avoid the risk of us exploiting inadvertent features of a model which may not be realised in practice.

\medskip

Our main finding is to quantify by how much collective drawdown outperforms the next-best alternative. The next-best in our model is a flex-and-fix strategy. It is outperformed by collective drawdown by a factor of 4.1\%, 13.1\% and 28.5\% in the Black--Scholes model, our SIR model and our richer model respectively. These are risk-adjusted comparisons.

The potential benefit over the worst-performing alternative is more than 30\%, 75\% and 90\% in our economic models.

Our numerical results also quantify the effect of the inefficiencies in shared-indexation designs that were identified in \cite{armstrong_dalby_donnelly}.

\medskip

Our second finding is to observe the importance of 
leverage in developing optimal investment strategies. For the representative investor that we consider, the optimal
strategies we find all take a leveraged position early
in life. Collective designs make this feasible without needing
complex financial engineering because the fund as a whole
is unlikely to take a highly leveraged position. This is because
the youngest investors hold a relatively small proportion of the assets in a scheme.

The next-best alternative to collective drawdown, flex-and-fix, is able to provide good results in retirement only if leverage is allowed. The optimal strategy in retirement is to annuitise most of one's wealth and to take a leveraged position with the remaining investments. The net position does not have a particularly high equity exposure, so this is not an exceptionally risky strategy, but it would be challenging to achieve in practice. This is due to both costs and the difficulty of achieving regulatory approval.

\medskip

The focus of the paper is on numerical results, but we also state a simple theoretical result which shows that, in the large fund limit, the longevity credits members obtain are independent of the investment decisions of other members. The possibility of decoupling investment risk and longevity risk was observed in \cite{donnelly2014} and \cite{armstrong_buescu_dalby}, but we describe an effective discrete-time design which gives convergence without any heterogeneity requirements.

\medskip

The structure of the paper is as follows.

In Section \ref{sec:cdc}, we describe the operation of CDC funds. The description of the collective drawdown design in Section \ref{sec:collectiveDrawdown} summarizes the design suggested in \cite{armstrong_buescu_dalby}. The description of shared-indexation designs in Section \ref{sec:shared_indexation} summarizes that of \cite{armstrong_dalby_donnelly}.

In Section \ref{sec:model}, we summarize how we have modelled the key risk factors in each of the economic models. The richer economic model we have chosen is the standard model used by the UK Pension Policy Institute (PPI). We have used it to facilitate comparison with other PPI reports. Details of this model are given in  \cite{armstrongMaffraPennanen}. The key additional detail described in this section is how we have extended the model to allow us to consider leveraged investments.

In Section \ref{sec:results}, we present our main numerical results comparing the performance of different scheme designs in each economic model for a chosen representative gain function.

\section{Collective scheme designs}
\label{sec:cdc}
We now describe the key features of the two CDC architectures modelled in this paper.

\subsection{Collective drawdown}
\label{sec:collectiveDrawdown}

The key feature of collective drawdown is an internal insurance market in idiosyncratic longevity risk. This internal market operates as a form of tontine,
where individuals agree that the funds of deceased individuals should be shared
among survivors. There is an extensive literature on tontines: see for example 
\cite{milevsky, milevsky2015, milevsky2016, chen, chenAndRach,chenRachSehner2020,
donnelly2014, donelly_bernhardt, boadoPenas} among many others.
Our term {\em collective drawdown} is intended to refer to the totality of a scheme design which includes both a selected investment and consumption strategy and a tontine mechanism for longevity pooling. The precise longevity pooling mechanism we will use was proposed in \cite{armstrong_buescu_dalby} and is very similar to the continuous-time mechanism proposed in \cite{donnelly2014}. The key feature of these approaches to longevity pooling is that they allow longevity risk to be pooled across members with different investment strategies.

\subsection{The internal insurance market}

Let us suppose that there are $n$ individuals whose probability of surviving over the period $[t,t+1]$ is $p^i \in (0,1)$ ($1 \leq i \leq n)$.  At the end of an investment period, suppose that the market value of each individual's fund is $v^i$. Let $S^i$ be an indicator taking the
value $1$ if individual $i$ survives over the period. The total
funds available to the longevity pool will then be given by
\[
F = \sum_{j=1}^{n}  (1-S^j) v^j.
\]
These funds are then shared among survivors in proportion to
their expected contribution once $v^i$ is known, which is defined to equal the odds of dying times their final wealth\begin{footnote}{There is an error in the description of the algorithm in \cite{armstrong_buescu_dalby} as it defines the contribution using the probability rather than odds.}\end{footnote}. So,
the amount received by individual $i$ will be
\begin{equation}
P^i_n:=
\frac{F S^i \frac{1-p^i}{p^i} v^i}{ \left(\sum_{j=1}^{n}  S^j\frac{1- p^j}{p^j} v^j\right)}
.
\label{eq:longevityCredit}
\end{equation}

We now wish to show that, subject to a bound on the mean initial wealth, in the large fund limit the value of this longevity credit does not depend upon the investment decisions of other members. Further, that it is equal to the longevity credit that would be received by an individual who invested in a classical tontine where all members have identical survival probabilities and investment portfolios. This is the substance of the next proposition.

\begin{lemma}
\label{lemma:tontineConvergence}
Let $\omega^M$ be a random variable representing market outcomes. Let $(t^i)_{i=1}^\infty$ be a sequence of independent identically distributed random variables denoting the ``type'' of individual $i$, and suppose $p^i=p(t^i)$ and $v^i=v(t^i,\omega^M)$ for some measurable functions $p$ and $v$ taking values in $(0,1)$ and $(0,\infty)$ respectively. Write ${\cal F}^{t^i}$ for the sigma algebra generated by $t^i$. Let $U^i$ be a sequence of i.i.d uniform random variables on $[0,1]$.
Suppose that $\omega^M$, $(t^i)_{i=1}^\infty$, and $(U^i)_{i=1}^\infty$ are independent.
Let ${\cal F}^M$ be the sigma algebra generated by $\omega^M$.
Let ${\cal F}^{M,i}$ be the sigma algebra generated by $(\omega^M,t^i,U^i)$.
Define 
\begin{equation}
S^i = \begin{cases}
0 & U^i > p^i \\
1 & \text{otherwise}.
\end{cases}
\label{eq:bernoulliCondition}
\end{equation}

Suppose ${\mathbb Q}$ is a pricing measure equivalent to the physical measure ${\mathbb P}$ with $\frac{d{\mathbb P}}{d{\mathbb Q}}$ being ${\cal F}^M$ measurable. Suppose
\begin{equation}
{\mathbb E}_{\mathbb Q}( v^j )<\infty.
\label{eq:momentBound}
\end{equation}
Then, defining $P^i_n$ using equation \eqref{eq:longevityCredit}, we have
\begin{equation}
\lim_{n \to \infty}
{\mathbb P}(P^i_n\leq x \mid {\cal F}^{M,i}) = 
{\mathbb P}\left(S^i\frac{1-p^i}{p^i} v^i\leq x \mid {\cal F}^{M,i}
\right).
\label{eq:conditionalConvergenceInDistribution}
\end{equation}
\end{lemma}
\begin{proof}
Equation \eqref{eq:conditionalConvergenceInDistribution}
is a statement on the conditional convergence in distribution of $P^i$. 

Since $\frac{d{\mathbb P}}{d{\mathbb Q}}$ is ${\cal F}^M$ measurable equation 
\eqref{eq:conditionalConvergenceInDistribution} is equivalent to the condition
\begin{equation}
\lim_{n \to \infty}
{\mathbb Q}(P^i_n\leq x \mid {\cal F}^{M,i}) = 
{\mathbb Q}\left(S^i\frac{1-p^i}{p^i} v^i\leq x \mid {\cal F}^{M,i}
\right),
\end{equation}
and so we may assume without loss of generality that ${\mathbb P}={\mathbb Q}$.

Furthermore, we assume without loss of generality that $i=1$ and rewrite equation \eqref{eq:longevityCredit} as
\begin{equation}
P^1_n= 
\frac{ \left(\frac{1}{n-1} (1-S^1) v^1 + \frac{1}{n-1} \sum_{j=2}^n (1-S^j) v^j \right) S^1 \frac{1-p^1}{p^1} v^1}{ 
\frac{1}{n-1}S^1\frac{1- p^1}{p^1} v^1
+
\frac{1}{n-1}
\left(\sum_{j=2}^{n}  S^j\frac{1- p^j}{p^j} v^j\right)}
.
\label{eq:expanded}
\end{equation}
Consider the random vector obtained from the summands in this expression:
\[
V^j=\left( (1-S^j) v^j, S^j \frac{1-p^j}{p^j}v^j\right).
\]
Its distribution is independent of the choice of $j \geq 2$. Let $\mathcal{G} := \mathcal{F}^{t^j} \vee \mathcal{F}^{1, M}$. Then by the tower property, followed by independence, we have
\[
\begin{split}
{\mathbb E}\left(S^j\frac{1-p^j}{p^j}v^j \mid {\cal F}^{1,M}\right)
&=
{\mathbb E}\left({\mathbb E}\left(S^j\frac{1-p^j}{p^j}v^j \mid {\cal G} \right) \mid {\cal F}^{1,M}\right) \\
&=
{\mathbb E}\left(\frac{1-p^j}{p^j}v^j{\mathbb E}\left(S^j \mid {\cal G} \right) \mid {\cal F}^{1,M}\right) \\
&=
{\mathbb E}\left((1-p^j)v^j \mid {\cal F}^{1,M}\right).
\end{split}
\]
So we see that
\[
\mathbb{E}(|V^j| \mid {\cal    F}^{1,M})
\leq 2 \mathbb{E}\left( \left|
  v^j \right| \mid {\cal F}^{1,M} \right)< \infty.
\]
The $V^j$ are conditionally independent by our independence assumptions.
Hence, by the conditional strong law of large numbers, $\frac{1}{n-1}\sum_{i=2}^n V^j$ converges in conditional distribution to its mean
\[
\begin{split}
{\mathbb E}\left(V^j \mid {\cal F}^{M,1}\right)
&= 
\left( \mathbb{E}((1-p^j) v^j \mid {\cal F}^{M,1}), \mathbb{E}((1-p^j) v^j \mid {\cal F}^{M,1})\right).
\end{split}
\]

The result now follows by applying the conditional continuous mapping theorem
to equation \eqref{eq:expanded} to compute the limiting conditional distribution of $P^1_n$.
\end{proof}

A proof of a similar result is given in \cite{donnelly2014}, but subject to the assumption that individuals can be grouped together into a finite set of types.  

In fact, no heterogeneity is required at all to define an effective tontine.
Choose any increasing sequence $(B_\alpha)_{\alpha=1}^\infty$ of reals with $B_1=0$ and $B_\alpha\to\infty$ as $\alpha\to\infty$, and
define $v^i_\alpha = \min\{ \max\{ v^i-B_\alpha,0 \}, B_{\alpha+1}-B_\alpha\}$. 
Define a new longevity credit value by
\begin{equation}
P^i_{B,n}:=
\sum_{\alpha=1}^\infty
\frac{(\sum_{j=1}^n (1-S^j)v^j_\alpha) S^i \frac{1-p^i}{p^i} v^i_\alpha}{ \left(\sum_{j=1}^{n}  S^j\frac{1- p^j}{p^j} v^j_\alpha\right)}
.
\label{eq:noAssumptionLongevityCredit}
\end{equation}
When the numerator is 0, we define the longevity credit to equal 0 even if the denominator is 0.

The next result shows that we obtain convergence in distribution for this tontine
design for all scheme members. To give a clean statement, we will write 
\[
X_n \underset{{n \to \infty}}{\overset{{d\mid{\cal F}}}{\to}} X
\]
if $X_n$ 
converges to $X$ in distribution conditioned on a sigma algebra ${\cal F}$.
\begin{theorem}
Let $\omega^M$, $(v^i)_{i=1}^\infty$,
$(p^i)_{i=1}^\infty$  and $(S^i)_{i=1}^\infty$ be as in Lemma \ref{lemma:tontineConvergence}, then
\begin{equation}
P^i_{B,n} \underset{{n \to \infty}}{\overset{{d\mid{\cal F}^{M,i}}}{\to}} 
S^i\frac{1-p^i}{p^i}v^i
\label{eq:convergenceInDistribution}
\end{equation}
almost surely even if the moment bound, equation \eqref{eq:momentBound}, does not hold.
\end{theorem}
\begin{proof}
Given $\epsilon>0$, choose $N$ such that $P(v^i>N)<\epsilon$. From Lemma \ref{lemma:tontineConvergence} we have convergence in distribution conditioned
on ${\cal F}^{i,M}$ when $v^i\leq N$ as only a finite number of terms in $\alpha$ will contribute to the longevity credit in  equation
\eqref{eq:noAssumptionLongevityCredit},
namely those with
$B_{\alpha}\leq N$. Here we are using the fact that the moment bound, equation \eqref{eq:momentBound}, holds for all $v^i_{\alpha}$. Hence the probability that 
the convergence in distribution \eqref{eq:convergenceInDistribution} occurs is at least $1-\epsilon$.
\end{proof}

In the simulations shown in the body of this paper we use a large fund limit when evaluating all scheme designs.
Appendix \ref{sec:finiteFund} shows
how the collective drawdown design performs when there are a small number of members in each generation. The paper \cite{armstrong_buescu_dalby}  contains numerical results which show the effectiveness of the tontine design when there are a small number of members.

\subsection{Shared-indexation}
\label{sec:shared_indexation}
\subsubsection{Informal discussion}

 In a shared-indexation design, all members' assets are pooled. Members all have a known ``nominal benefit amount'' which they have accrued to date, and each year all members receive the same percentage increase or decrease to their accrued nominal benefit amount to reflect the change in asset performance over the year. The level of increase or decrease is chosen so that, using a central-estimate methodology, the funds current assets match its projected outgoings assuming the same level of indexation is applied each year. The level of indexation currently applied is called the {\em prevailing level of indexation}. This is the behaviour of the scheme in a typical year, but there are additional mechanisms to apply cuts and bonuses if the prevailing level of indexation falls outside a given range. The shared prevailing level of indexation is what gives this design its name.

After retirement, members receive their nominal benefit amount each year. However, before retirement, the nominal benefit amount is best thought of simply as an accounting device. It is not a projection of expected benefits.

In addition to this shared-indexation mechanism, a scheme must specify how much members are charged
in order to accrue additional nominal benefit entitlement. In a {\em flat-accrual} design, members receive additional nominal benefits in proportion to their contribution regardless of their age or fund performance. This is the approach taken in the Royal Mail Collective Pension Plan. The present value of the additional pension income for a given level of contribution will be much larger for older members than younger members as the flat-accrual design does not account for the time-value of money. This mirrors the design of CARE defined-benefit (DB) schemes and is a deliberate design feature. The flat-accrual design is intended as a replacement for existing DB designs so it seeks to replicate this aspect of their behaviour.

As an alternative to flat-accrual, a shared-indexation design may follow what we call a {\em dynamic-accrual design}.  In this design the benefit received for a given level of contribution varies with both age and the prevailing level of indexation. The design requires benefits to match contribution when benefits are priced using a central-estimate methodology on the assumption that indexation will remain constant. The intention is that this design should reduce inter-age cross-subsidies. In 2024 the Department for Work and Pensions (DWP) consulted on multi-employer CDC designs, and the design envisaged by this consultation was a shared-indexation design but with dynamic-accrual.

\medskip


\subsubsection{Mathematical formalisation}

The following description summarizes that given in \cite{armstrong_dalby_donnelly}.

We assume the scheme contains $M$ different generations indexed by $\xi\in\{0,1\ldots,M-1\}$. Individuals within each generation are identical apart from their individual mortality risk.

Each year, all accrued pensions are increased by the same pension increase factor $\theta_t (1 + h^{\nom}_{t})$, in which $\theta_t$ is a \emph{bonus level} and $h^{\nom}_{t}$ is the {\em nominal indexation rate}. Denote by $\DCPIt=\frac{\CPI_t}{\CPI_{t-1}}-1$ the annual effective rate of change in the Consumer Price Index, $\CPI_t$, from time $t-1$ to time $t$.  Define the {\em real indexation rate} $h_t$, at time $t$, as the solution to
\[
1 + h^{\nom}_{t} = (1 + h_t) \left( 1 + \DCPIt \right).
\] 

We define the \emph{nominal benefit} for each member at the start of the next period, before any further
contributions are made as
\begin{equation}
B^{\xi,\cum}_{t-} := \theta_t \, (1 + \DCPIt)(1 + h_t) \, B^{\xi,\cum}_{t-1}, \quad \text{for }t\geq 1,
\label{eq:cdcDefining}
\end{equation}
where $B^{\xi, \text{cum}}_\tau=0$, if $\tau$ is the year a member joins the scheme. That is, members do not accrue benefits until they begin contributing to the scheme.

It is convenient to make the following definitions:
\begin{itemize}
\item $N^{\xi}_t$ be the number of surviving individuals of type $\xi$ in the fund at time $t$.

\item $p^{\xi}_{t,k}$ be the probability that an individual of type $\xi$ survives from time $t$ to time $t+k$, given that they are alive at time $t$.  It is assumed that future lifetimes are independent random variables.

\item The fund as a whole invests in a pre-determined
asset mix at each time $t$. Let $\nu_{t,k}$ be the cumulative discount factor that discounts 1 unit paid at time $t+k$ back to time $t$.  Let $\hat{R}_{t,k}$ be the annual effective rate of return predicted at time $t$ of the assets invested, in the time period $[\max\{t+k-1,t\}, t+k)$. Then, for $k \in \mathbb{N}_0$,
\[
\nu_{t,k} = \prod_{n=0}^{k} (1+\hat{R}_{t,n})^{-1}.
\]
Note that $\hat{R}_{t,0}=0$.

\item $\widehat{\DCPI}_{t,k}$ be the projection made at time $t$, of the annual change in CPI from time $t+k-1$ to time $t + k$, for $k\in\mathbb{N}_0$.  Note that $\widehat{\DCPI}_{t,0}=\DCPIt$, the actual change in CPI over the year to time $t$.
\item $I_{t,k}(\aitch)$ be the projection made at time $t$ of the cumulative nominal indexation rate from time $t$ to time $t+k$, using a constant real indexation rate $\aitch$, for $k \in \mathbb{N}_0$, which is calculated as
\[ 
I_{t,k}(\aitch) := \prod_{n=0}^{k} (1 + \widehat{\DCPI}_{t,n})(1 + \aitch).
\]

\item ${\mathbf 1}^{\xi, \Ret}_t$ be the deterministic indicator function taking the value $1$ if individuals of type $\xi$ have retired at time $t$, and $0$ otherwise.

\item $A_{t-}$ be the value of assets at time $t$, just before new contributions are added or benefit payments are made at time $t$.

\item $\pi_t$ be the proportion invested in risky assets in year $t$.
\end{itemize}

The values of $h_t$ and $\theta_t$ in years $t>0$, are determined by the total asset value at time $t$ before payments in and out of the scheme, which we denote by $A_{t-}$. We then
require that $h_t$ and $\theta_t$ satisfy the equation
\begin{equation}
A_{t-} = \theta_t \sum_{\xi=0}^{M-1} N^\xi_t \, B^{\xi,\cum}_{t-1} \, \sum_{k=0}^\infty p^{\xi}_{t,k} \, \nu_{t,k} \, I_{t,k}(h) \, {\mathbf 1}^{\xi, \text{Ret}}_{t+k}.
\label{eq:actuarialvaln}
\end{equation}
If it is possible to solve this equation with $h_t \in [h_{\min},h_{\max}]$ and $\theta_t=1$, then these are the values selected. Otherwise, we take $h_t=h_{\min}$ or $h_t=h_{\max}$ and then choose $\theta_t<1$ or $\theta_t>1$ respectively, as to solve this equation. In the latter cases, a benefit cut or bonus is said to have occurred.

The additional benefit entitlement $B^\xi_t$ a member receives in exchange for a contribution $C^\xi_t$ depends upon the specific type of shared-indexation design.

\begin{enumerate}[(i)]
\item 
In a {\em dynamic-accrual} design, the additional benefit entitlement, $B^\xi_t$, received by generation $\xi$ in exchange for a contribution, $C^\xi_t$, satisfies:
\begin{equation}
\label{eq:multiemployerBenefit}
C^\xi_t = \frac{B_t^\xi}{I_{t,0}(h_t)} \, \sum_{k=1}^\infty I_{t,k}(h_t) \, \nu_{t,k} \, p^\xi_{t,k} \, {\mathbf 1}^{\xi, \text{Ret}}_{t+k}.
\end{equation}
\item
In a {\em flat-accrual} design
\[
B^\xi_t=\alpha C^\xi_t
\]
for some constant chosen by the scheme designer. The value of $\alpha$ is chosen so that when the prevailing level of indexation matches a chosen target level and
if asset returns are as expected equation \eqref{eq:cdcDefining} will
continue to hold at the start of the next period.
\end{enumerate}
The new cumulative benefit entitlement for each generation is
\[
B^{\xi,\cum}_t = B^{\xi,\cum}_{t-}+B^\xi_t.
\]
The pension received by each generation is equal to
\[
B^{\xi,\cum}_t {\mathbf 1}^{\xi,\Ret}_{t}.
\]

The fund asset value after benefit payments and contributions is given by
\begin{equation}
A_{t} := A_{t-} + \sum_{\xi=0}^{M-1} N^\xi_t (C^\xi_t - B^{\xi,\cum}_t {\mathbf 1}^{\xi,\Ret}_t).
\label{eq:increments}
\end{equation}
The scheme's asset value evolves as
\[
A_{t-}=\begin{cases}
0 & \textrm{if $t = 0$}, \\
(1 + R_t) A_{t-1} & \textrm{if $t=1,2,\ldots$}.
\end{cases}
\]
Here, $R_t$ denotes the return realised by the fund on its investments over the time period $(t-1,t]$.

In a flat-accrual scheme, the proportion invested in risky assets is determined
by choosing a lifestyle strategy for each investor. This specifies the desired proportion to invest in risky assets given the age of the investor. The overall position taken by the fund is a liability-weighted average of these proportions for each investor. In practice, the details of the lifestyle strategy have their most significant impact during startup and shutdown of the scheme and will tend to be approximately constant when the scheme is in a steady state.

To choose the proportion invested in risky-assets in a dynamic-accrual scheme, we first simulate a flat-accrual scheme and record the median proportion invested in risky assets each year. When simulating the dynamic-accrual scheme, we then use these proportions. This ensures that the dynamic-accrual scheme behaves similarly to the flat-accrual scheme during startup and shutdown.

A shared-indexation fund can now be simulated using the procedure defined by these equations.

\medskip

The motivation for the shared-indexation design is somewhat obscure. The paper \cite{armstrong_dalby_donnelly} provides some motivation and describes the historical and legislative context which led to these designs. Nevertheless, it is important
to realise that these designs are not derived from any single
economic optimisation principle, but have emerged instead
as a developing consensus among UK pension professionals on
how collective schemes might be developed.

The practical consequence of the design is that generations who are closer to retirement will experience changes to indexation for a shorter time, members who are closer to retirement will experience less volatility in the market value of their future benefits than members who are a long way from retirement. This smooths incomes in retirement while providing leverage for younger members. As we shall see later, the optimal investment strategies are qualitatively similar in this regard.

The different designs of shared-indexation scheme are studied in detail in
\cite{armstrong_dalby_donnelly} and we will now summarize the key findings of that paper.

The flat-accrual design is intended to mimic the benefit structure of a CARE DB scheme.
As a result it contains significant inter-age cross-subsidies by design, but these are in fact much larger than the cross-subsidies in CARE DB schemes. These large
cross-subsidies can result in an effect termed {\em drag} where early generations receive a pension worth more than they pay and subsequent generations receive a
pension worth less than they pay in.

The dynamic-accrual design is intended to price benefits fairly, but the pricing formula used assumes indexation will remain constant. This counterfactual assumption leads to pricing errors which can be detected by pricing benefits by Monte Carlo in a risk-neutral model.


\section{Modelling details}\label{sec:details}
\label{sec:model}
\subsection{The Black--Scholes economic model}
In this model, the simplest of all three models, there exists only one risky asset, with the ability to trade simultaneously in a risk-free rate. The dynamics of the risky asset price, denoted $S$, follow a geometric Brownian motion described by
\begin{equation*}
    \ed S_t = \mu S_t \ed t + \sigma S_t \, \ed W_t, \label{eq:BSM}
\end{equation*}
where $\mu \in \mathbb{R}$ represents the drift, $\sigma \in \mathbb{R}_{+}$ the volatility and $W$ a standard Brownian motion. Investment and consumption decisions are assumed to be made at discrete intervals, defined by the set ${\cal T}:=\{0,\delta t, 2 \delta t, \ldots, T \}$ for some time-step $\delta t$ and final time $T$. Between these times, investments are made following a fixed-weight strategy. Let $\pi_t$ be the proportion of wealth allocated to the risky asset at time $t \in {\cal T}$. This proportion is fixed throughout the interval $[t,t+\delta t)$ with the remainder of the portfolio allocated to a risk-free asset, growing at a constant rate $r$.
We denote the limit from the left of wealth at the end of the period by $w_{(t+\delta t)-}$.
Applying It\^o's lemma and solving the resulting linear SDE
yields
\begin{equation*}
    \log(w_{(t+\delta t)-}) - \log(w_t) = \left(\pi_t \mu + (1-\pi_t)r-\frac{1}{2}(\pi_t\sigma)^2 \right) \delta t + \pi_t\sigma (W_{t+\delta t}-W_t).
\end{equation*}
This yields the left limit for $w_{(t + \delta t)-}$. We define
\begin{equation}
\epsilon_t := \frac{W_{t+\delta t}-W_t}{\sqrt{\delta t}},
\label{eq:epsilon_as_normal}
\end{equation}
so $\epsilon_t$ is distributed according to standard normal. Thus we are able to simulate the log wealth process using the Gaussian increments $\epsilon_t$, and  this log simulation, combined with the fixed-weight strategies, automatically ensures that strategies that put one into debt are removed. 

\subsection{Stochastic Interest Rate (SIR) model}

In this economic model, we have two assets: the first is an equity index which follows a real-terms geometric Brownian motion, the second is a long-term bond index whose yields are assumed to follow mean-reverting dynamics. We do not allow investment in short-term bonds, so this is an incomplete market model.

Because of this incompleteness, it will not be possible to replicate bonds with arbitrary maturities. If we were to find the optimal investment in a complete market model, we would be implicitly assuming that our model accurately reflects the dynamics of the both the underlying and derivatives contracts. This is unrealistic. By restricting the asset classes which can be used for investment, we do not need to find a model which can be calibrated to the entire yield curve. 

The only bond portfolio in which we will allow investment is the index-linked bond portfolio required to perfectly hedge a fund's liabilities. We will work in an index-linked numeraire so that the equity index will follow geometric Brownian motion and inflation will be constant and equal to zero. This results in a more parsimonious model with only one risk factor determining the yields of our bond index rather than separate risk factors for interest rates and inflation. If we were to allow both index-linked and non-index-linked investments, we would need to introduce two risk factors.

For collective drawdown and individual investment designs, the bond portfolio required to perfectly hedge liabilities is the bond portfolio required to hedge a single-life annuity. The duration of this portfolio will vary with age, which will become shorter over the lifetime. We will refer to the situation where the duration varies throughout the simulation as case (i).

For the shared-indexation design, during the steady state of the scheme the bond portfolio required to perfectly hedge liabilities will have an approximately constant duration. We will call this case (ii).

To accommodate these two different situations, we will assume that the yield for our index can be computed by using a discrete-time short-rate model and then computing the yield of either i) the bond portfolio that hedges an annuity or ii) a bond portfolio of fixed duration. This will yield comparable dynamics for the bond indices across the two models. This differs from assuming a discrete-time short-rate model in that we do not assume that short-dated bonds follow the dynamics determined by this short-rate model. Indeed, we do not even assume that short-dated bonds are tradable. This is reflected in the calibration process for our model: we calibrate the parameters for our short-rate model by fitting it to the historical data for a long-term bond index.

In summary, we use a short-rate model to generate dynamics for a long-term bond index. Because the long-term index is the only one in which we can trade, we must calibrate to this long-term index. Because we do not allow trading in other bonds, it is not a concern that the yield curve that would arise from our short-rate model were such trading allowed may be very different from the realised yield curve. The bond index we will use is different in case (i) and case (ii), so these are not identical market models but the price of the tradable assets in both arise from a common short-rate model and so in this sense are mutually consistent.

\medskip



Let us now describe the model more precisely. The short rate is modelled by a discrete-time Ornstein--Uhlenbeck process:
\[
r_{t+\delta t} = r_t + a(b-r_t)\delta t + c\sqrt{\delta t}\,\xi_t,
\]
where $\xi_t$ are i.i.d.\ standard normal random variables, and their correlation with the noise driving the risky asset is given by $\rho$. Our choice of a discrete-time short-rate model is deliberate, and reflects how interest rates may jump at a fixed time in response to announcements or new policy from government bodies or central banks. 

By fixing a time $t$, we determine the bond portfolio we will use for computation from time $t$ to $t + \delta t$, which has a unit price denoted by $B_t$. Let us consider case (i) where this portfolio perfectly hedges an annuity. The price of this portfolio at time $t$ is given by:
\begin{equation*}
    B_t = \sum_{s=\max\{t + \delta t,t_{RA}\}}^{T} p_{t,s} P(t, s),
\end{equation*}
where $p_{t,s}$ is the probability of surviving to time $s$ conditioned on being alive at time $t$ and $P(t, s)$ is the price at time $t$ of a zero coupon bond maturing at time $s$. 

We can use this to compute the yield of the portfolio, by solving for $y^{B_t}$ in 
\begin{equation}
    B_t = \sum_{s=\max\{t + \delta t,t_{RA}\}}^{T} p_{t,s}\exp\bigl(-y^{B_t}(s-t)\bigr).
    \label{eq:yield}
\end{equation}
We assume that $y^{B_s}$ remains constant for $s \in [t, t+\delta t)$. So, the bond price
grows deterministically at rate $y^{B_t}$, yielding the left limit
\begin{equation}
B_{(t+\delta t)-} = B_t e^{y^{B_t} \delta t}.
\label{eq:bondGrowth}
\end{equation}

Computing the yield \(y^{B_t}\) of the bond portfolio requires solving the nonlinear equation in \eqref{eq:yield}, which we can do numerically using a root-finding algorithm. Repeating this computation at every time step and for every simulated short-rate path is computationally expensive. To reduce this computational cost, we approximate the mapping from the short rate \(r_t\) to the corresponding bond portfolio yield \(y^{B_t}\) using a polynomial. 

Let $\mathcal{R} = \{ r_t^{(k)} : t = 0,\dots,T,\; k = 1,\dots,K \}$ denote a grid of short-rate values where $t$ denotes the time component and $k$ denotes the spatial component. For each grid point, we compute the corresponding bond portfolio price
\begin{equation*}
    B_t\bigl(r_t^{(k)}\bigr)
    =
    \sum_{s=\max\{t+\delta t,\,t_{RA}\}}^{T}
    p_{t, s} P\bigl(t,s; r_t^{(k)}\bigr).
\end{equation*}
By substituting these prices into equation~\eqref{eq:yield}, we obtain the associated portfolio yields
$y^{B_t}(r_t^{(k)})$, by using the root solver at each grid point. We then fit an $n$ degree polynomial approximation, using least squares, to the mapping
$r_t \mapsto y^{B_t}(r_t)$ over the grid $\mathcal{R}$. This yields time-dependent coefficients
$\{C_{i,t}\}_{i=0}^n$ such that
\begin{equation*}
    y^{B_t}(r_t)
    \approx
    \sum_{i=0}^n C_{i,t} r_t^i.
\end{equation*}
The coefficients depend on time since the cash flows of the bond portfolio vary with the age of the individual. During simulation, whenever the short rate $r_t$ is observed, the portfolio yield is approximated by evaluating the fitted polynomial. Using a degree 10 polynomial approximation allows rapid calculations of the bond prices with negligible numerical error.

Although changes in the yield of our fixed income portfolio  will occur in discrete time, we use a continuous-time trading model in order to model leveraged investments. Let $w_t$ denote total wealth and $\pi_t$ the fraction invested in the stock. The remainder is invested in the bond portfolio which grows according to equation \eqref{eq:bondGrowth}.
We assume $\pi_t$ is chosen at time $t$ and held constant on $[t,t+\delta t)$. Holding $\pi_t$ fixed on $[t,t+\delta t)$, we write the SDE for the log wealth as
\begin{equation}
d\log w_s
=
\Bigl(\pi_t\mu + (1-\pi_t)y^{B_t} - \tfrac12\pi_t^2\sigma^2\Bigr)\,ds
+
\pi_t\sigma\,dW_s, \quad s \in [t, t+ \delta t).
\label{eq:ctsTimeLogWealthSIR}
\end{equation}

We define $w_{(t+\delta t)-}^\dagger=\lim_{s\to t+\delta t} w_s$.
We can simulate this in one step using Gaussian increments. It represents the wealth before any jump in interest rates. 
At the end of the period the short rate jumps from \( r_t \) to \( r_{t+\delta t} \), inducing a corresponding jump in the bond price. Accounting for this, we define
\begin{equation*}
    w_{(t+\delta t)-}
    =
    w_{(t+\delta t)-}^\dagger
    \left(
        1
        +
        \frac{1 - \pi_{(t+\delta t)-}}{B_{(t+\delta t)-}(r_t)}
        \bigl[
            B_{(t+\delta t)-}(r_{t + \delta t})
            -
            B_{(t+\delta t)-}(r_t)
        \bigr]
    \right).
\end{equation*}
The quantity $\pi_{(t+\delta t)-}$ represents the portfolio weight at the end of the period and we ensure admissibility of strategies by constraining it to  lie in the interval \([0,1]\):
\[
\pi_{(t+\delta t)-} =
\begin{cases}
1 & \text{if } \pi_t \geq 1, \\
\pi_t & \text{if } 0 < \pi_t < 1, \\
0 & \text{if } \pi_t \leq 0.
\end{cases}
\]
This guarantees that wealth at the end of the investment period, $w_{(t+\delta t)-}$ is non-negative. Since interest-rates jump at known times in our model, this is an admissible strategy.

The modelling described above ensures that we have complete consistency between the price process for our bond index and the price of annuities. This is essential when modelling the flex-and-fix strategy which allows dynamic trading in both bonds and annuities.

When modelling a shared-indexation scheme, there is no trading allowed in annuities, only investment in a bond index. As a result, we do not need such a complex model in case (ii). In this case, the fund wishes to hedge liabilities seen on the right hand side of equation \eqref{eq:actuarialvaln} rather than a single-life annuity. We therefore assume that bond portfolio used for investment each period will match the payment structure of equation \eqref{eq:actuarialvaln} and grows at a deterministic rate, $y_t$, over the investment period. Calculating this yield exactly as we have done for annuities would be computationally expensive. Instead we assume that the yields of the complex bond portfolio are the same as the yields of an index-linked zero coupon bond in our short-rate model with the same duration. The accuracy of this duration-based approximation of bond indices is shown in \cite{koivu2005modeling}, so this will yield a comparable model to that of case (i).

The model in this case (ii) is the same as the SIR model of \cite{armstrong_dalby_donnelly}. That paper describes how the parameters were calibrated, and we use the same parameter values for both cases.

\subsection{A richer economic model}

The SIR model of the previous section was developed as a parsimonious tool specifically for this research project. However, it is reasonable to ask how the different designs perform if one uses pre-existing models.
In the next model we will consider, all risk factors are simulated using a minor variation of the economic scenario generator (ESG) described in \cite{armstrongMaffraPennanen}, and which was calibrated independently from this paper.
This model incorporates features such as wage inflation and cyclical economic fluctuations which are not present in the SIR model.

At the heart of
this model is a vector-autoregressive model for a vector of risk-factors ${\mathbf{x}}$ at each time. This model takes the form
\begin{equation}
    \mathbf{x}_{t+1} - \mathbf{x}_t = A\mathbf{x}_t + \mathbf{b} + L\mathbf{\epsilon},
\label{eq:discreteVAR}     
\end{equation}
where $A\in \mathbb{R}^{n\times n}$ is the autoregression matrix, $b\in\mathbb{R}^n$ is a vector that allows one to set views about the long-term values of the risk factors, $\epsilon$ is a multivariate normal random variable with zero mean and $L\in\mathbb{R}^{n \times n}$ is the Cholesky decomposition of the associated covariance matrix $\Sigma\in\mathbb{R}^{n\times n}$.

The components of the vector $\mathbf{x}_t$ are derived from a sequence of transformations applied to economically relevant variables. We will not list all the variables in the model, but the Table \ref{table:varSummary} lists the formulae used to calculate the five components of $\mathbf{x}_t$ that are particularly relevant to this paper. In this table: $\SI_t$ denotes the value of a total return stock index at time $t$; $\CPI_t$ denotes the value of the consumer price index; $\YTM_t^\ell$ denotes the yield to maturity of a representative long-term bond index and $\LTIE_t$ denotes the long-term inflation expectation (defined as the difference between long-term yields and long-term index-linked yields: $\LTIE_t = \YTM_t^\ell - \YTM_t^i$); $\AWE_t$ denotes average weekly earnings.

\begin{table}
\begin{center}
\begin{tabular}{ll}
Index & Formula \\
\hline
1  & $\log(\SI_t)-\log(\SI_{t-1})$ \\
2  & $\log(\CPI_t)-\log(\CPI_{t-1})$ \\
3  & $\log\left( \frac{\YTM^{\ell}_t}{\CPI_t/\CPI_{t-1}} + 0.05\right) $ \\
4  & $\LTIE_t - \log\left(\CPI_{t}/\CPI_{t-1}\right) $ \\
5  & $\log(\AWE_t)-\log(\CPI_t) $ \\
\end{tabular}
\caption{The five components of the risk-factor vector $\mathbf{x}_t$ most
relevant to this paper}
\label{table:varSummary}
\end{center}
\end{table}

In the scenario generator, we simulate the vector ${\mathbf x}_t$ and then reverse the transformations
to compute the current values of the various indices and yields. To compute the value
of an index-linked bond index, $P^i_t$, the model uses the approximation formula
\[
\frac{P^i_{t+\delta t}}{P^i_{t}} \approx \exp\left(\YTM^i_t (\delta t) - D^i (\YTM^i_{t+\delta t}-\YTM^i_t) + \log( \CPI_{t+\delta t}) - \log( \CPI_{t} )\right)
\]
where $D^i$ is the duration of the index-linked bond portfolio, which is assumed to be constant.

The resulting scenario generator allows one to input views about long-term median values for different risk-factors and these are chosen as shown in Table
\ref{table:riskFactors}.

Although the scenario generator also allows one to simulate 
population mortality, we do not use this feature and instead use the S1PMA tables produced by the Continuous Mortality Investigation. This is because it is unclear
how shared-indexation designs will adapt to changes in mortality so we wish
to remove this as a point of comparison between scheme designs.

To compute projected values of CPI to apply in equation 
\eqref{eq:actuarialvaln}, we use median values as these are
simple to project in our ESG. Our ESG is designed to output index-linked
long-term bond yields and we use these combined with the long-term
median estimate for CPI to determine the discount rate for riskless assets in equation \eqref{eq:actuarialvaln}. We also assumed
that the bond-portfolio is index-linked and chosen to match the liabilities exactly. The returns on the bond portfolio can then be computed by pricing this portfolio using the index-linked long-term bond yields supplied by the ESG. The net result of these assumptions is that if the fund is invested 100\% in bonds, then $h$ will remain constant unless the lower bound on indexation is hit due to changes in CPI.

In our ESG, the stock follows geometric Brownian motion, so we can compute
the predicted mean returns and use this to compute the appropriate
discount rate for risky assets in equation \eqref{eq:actuarialvaln}.

As in the SIR model, we wish to allow leveraged investments
by modelling the evolution of the log wealth as is done in equation \eqref{eq:ctsTimeLogWealthSIR}. To achieve this we need a continuous
time model. In appendix \ref{sec:ctsTimeESG} we describe how to identify a continuous time model compatible with the discrete time model of \cite{armstrongMaffraPennanen}.
We will use the resulting equation \eqref{eq:logWealth}
to compute the increment of the wealth each period.

This economic model does not provide a full term-structure for interest rate products. In our simulations we assume that the index-linked long-term bond yields provided by the model can be used to price annuities and to calculate the investment returns from the bond portfolios in both models. This yields an arbitrage free model for both case (i) and case (ii). The weakness in comparison to the handling of the term structure for the SIR model is that it does not take account of the changes in duration in case (i). On the other hand, the model does incorporate new features such as wage inflation and so it might be considered to give a more accurate assessment of the overall risk.

\subsection{Model parameters and assumptions}
In all three models, to obtain $w_{t}$ from $w_{t-}$, we incorporate contributions, consumption and longevity payments via the equation
\begin{equation}
    w_{t} = \eta s_t \mathbbm{1}_{t<t_{\RA}} + (1- c_t
    \mathbbm{1}_{t\geq t_{\RA}})(1+ P_{\infty,t} \mathbbm{1}_{t>t_{\RA}}  )w_{t-},
\label{eq:wealthInfinite}    
\end{equation}
where the first term describes the fraction, $\eta$, of an individuals salary, $s_t$, that is contributed before retirement ($t_{\RA}$ is the time of retirement) and the second term removes the consumption, $c_tw_{t-}$ with $c_t \in (0, 1)$, and adds on any longevity payment, $P_{\infty, t}w_{t-}$, that one may receive in retirement. Since all of the terms in this equation \eqref{eq:wealthInfinite} are positive, this relation is consistent with wealth being prevented from going negative at any point in time.

When the scheme is first opened
there are an equal number of employees at all ages between 25 and the retirement age
which we take to be 65. Each year an equal number of employees joins at age 25.
All members have the same salary, which grows with the average weekly earnings index.
The fund is closed to new members after $100$ years.

We use the IFA's S1PMA tables to model mortality, assuming that the times of death of members are independently distributed. We assume the fund size is large, so longevity risk can be perfectly hedged. 

The long-term economic assumptions and parameter values used in our simulations
are summarised in Table \ref{table:riskFactors} and have been taken from OBR figures.
The contribution rate we have chosen matches the auto-enrolment default for
DC pensions in the UK. The other parameters of our model are calibrated statistically. The calibration of the rich economic model is described in \cite{armstrongMaffraPennanen} and the calibration of the SIR model in \cite{armstrong_dalby_donnelly}.

\begin{table}[h!tbp]
\begin{center}
\begin{tabular}{ll|ll}
\multicolumn{2}{c|}{Common parameters} & \multicolumn{2}{c}{Shared-indexation} \\ \hline
Long-term stock growth & 7.73\% & $h_{\min}$ & $0\%$ \\ 
Stock volatility & 15.3\% & $h_{\max}$ & $\CPI + 5\%$  \\
Long-term wage growth & 3.83\% & M & 150  \\
Long-term CPI growth & 2.00\% & &   \\
Long-term index-linked bond growth & 4.46\% & & \\
Contribution rate & 8\% & &  
\end{tabular}
\end{center}
\caption{Parameter values used in our simulations.}
\label{table:riskFactors}
\end{table}

In our DC models featuring annuity purchase, annuities are priced using the long-term index-linked bond yield with a $5\%$ additional charge added to represent
the risk-margin that an insurer must add due to systematic longevity risk.

\subsection{Optimization}
\label{sec:loss}

For DC designs, and collective drawdown, it is clear
what it means to identify an optimal strategy once one has defined a loss function. All the resulting investment problems can all be seen as variations on the classical Merton problem.
As the stochastic model we use is much richer than Merton's there is little hope of solving the problem analytically, or even using classical finite-difference methods. Instead we solve the optimal investment problems using machine-learning techniques.

To describe our preferences, we assume that at each time $t$ after retirement at time $t_{\RA}$, the individual consumes an amount $C_t$. We define
a function
\[
u(C_t)=\frac{C_t^\rho}{\rho}-\frac{a^\rho}{\rho}
\]
where ${\rho}<1$ and $a>0$ are parameters whose interpretation we explain shortly.
We assume that each individual seeks to maximize a gain function of the following form,
\begin{equation*}
    U(c) = \mathbb{E}\left( -\exp\left(-\alpha \sum_{t=t_{\RA}}^{t<\tau} u(C_t) \delta t\right)\right).
\end{equation*}
The parameter $\alpha>0$ is a risk aversion parameter. The variable $\tau$ is a random variable denoting the time of death, which is assumed to always be greater than $t_\RA$. The size of a time step is $\delta t=1$. The parameter 
$\rho$ is a satiation parameter i.e., it determines the rate at which the individual becomes satiated due to consumption at a given time. We will call the parameter $a$ 
the {\em adequacy level}, because consumption below the adequacy level in a given year decreases utility.  We have omitted indices to indicate the dependence on the individual for ease of notation.

These preferences are a variation on the concept of Kihlstrom--Mirman preferences \cite{kihlstromMirman},
but additionally incorporate mortality. Kihlstrom--Mirman preferences are not commonly used
in general economic applications as they can lead to time-inconsistent preferences \cite{krepsPorteus,epsteinZin1}. However, this specific form of the preferences using an exponential and no discounting factor, is time-consistent (see \cite{epsteinZin1}).
We do not need to include a discount factor because incorporating mortality in
the preferences already provides an incentive for earlier consumption. We refer
to this preference model (with arbitrary $u$) as exponential Kihlstrom--Mirman (EKM) preferences. In \cite{bommier} Bommier argues from an axiomatic approach that preferences in problems featuring mortality should have this form.

Similar to Epstein--Zin preferences, our preference model includes separate
parameters determining risk and satiation. Existing research 
suggests that separate parameters are necessary to resolve observed asset pricing puzzles \cite{bansalYaron,bansal,benzoniEtAl,bhamraEtAl}. This strongly suggests that the preferences of members of our fund will also require separate parameters for risk and satiation. However, unlike Epstein--Zin preferences, our preferences do not require a recursive computation of the utility in terms of conditional expectations at each time point. This makes the gain function \eqref{eq:utility} much simpler to evaluate than Epstein--Zin preferences. As a result, our choice of
preferences is well-suited to a machine-learning approach which will require rapid
computation of this utility.

We obtained the optimal strategy for a collective drawdown scheme using a recurrent neural network. The network receives economic data at each time point as input, specifically, wage growth, inflation and asset returns. It then outputs the proportion of wealth to consume at each time and the proportion to invest in each of the assets. From this
one can compute the replacement ratio in retirement i.e., the inflation linked proportion of final salary received. The replacement ratio at each time is fed into a loss function which scores the outcome. The network minimises this function. In  \Cref{sec:lossComputation} we describe how the loss
function is computed in practice and in Section \Cref{appendix:neuralNetworkDesign} we describe our neural network architecture.

\medskip

\begin{table}[h!tbp]
\begin{center}
\begin{tabular}{c|c|c}
\multicolumn{1}{c}{} & \multicolumn{2}{c}{} \\ \hline
\textbf{Loss Parameters} & \textbf{Black--Scholes \& SIR} & \textbf{Richer Economic Model} \\ \hline
$\alpha$ & $5.0 \times 10^{-5}$ & $2.5 \times 10^{-5}$ \\ \hline
$\rho$ & \multicolumn{2}{c}{-2} \\ \hline
$a$ & \multicolumn{2}{c}{0.4}
\end{tabular}
\end{center}
\caption{Parameters of our loss function.}
\label{table:lossParams}
\end{table}
The parameter values we chose for our loss function are shown
in Table \ref{table:lossParams}.
To select these parameters, the process we followed was to 
first simulate a dynamic-accrual shared-indexation scheme that matched
industry expectations. We then identified
a gain function such that the collective drawdown design gave comparable outcomes.

\begin{figure}[htp!]
  \centering

  \begin{subfigure}[b]{0.6\textwidth}
    \centering
    \includegraphics[width=\linewidth]{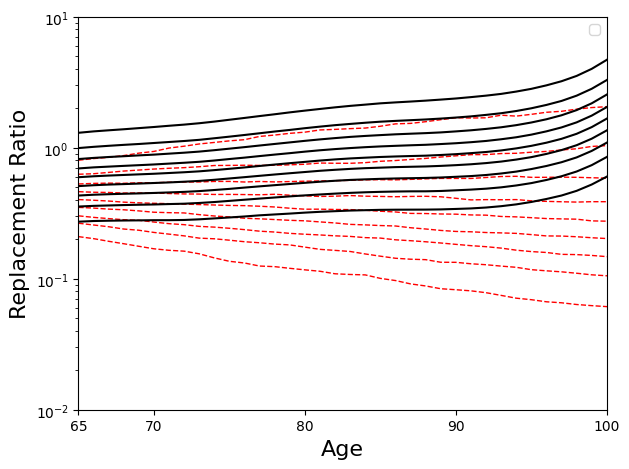}
    \caption{}
    \label{fig:black-scholes-RRs}
  \end{subfigure}
    \hfill
  \begin{subfigure}[b]{0.6\textwidth}
    \centering
    \includegraphics[width=\linewidth]{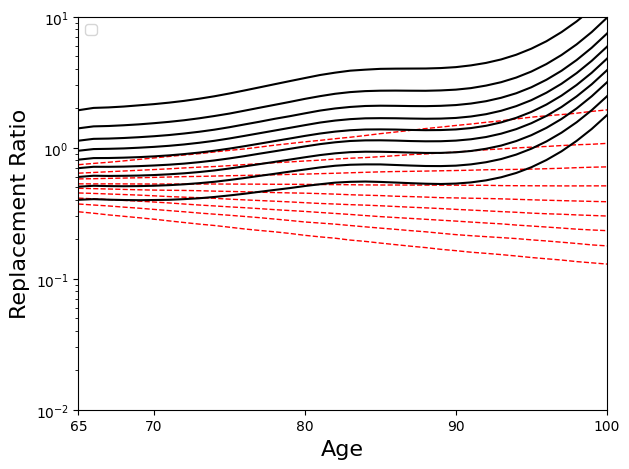}
    \caption{}
    \label{fig:SIR-RRs}
  \end{subfigure}

  \begin{subfigure}[b]{0.6\textwidth}
    \centering
    \includegraphics[width=\linewidth]{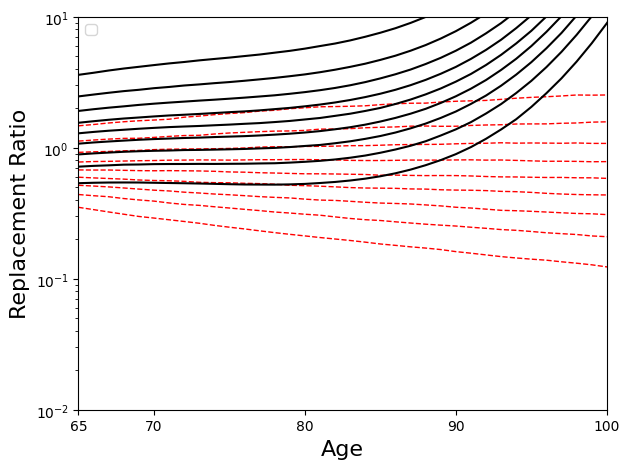}
    \caption{}
    \label{fig:stochasticDomination}
  \end{subfigure}

  \caption{The log replacement ratio of collective-drawdown (solid, black line) vs a shared-indexation CDC (dashed, red line) calibrated to industry expectations in the Black-Scholes model (\subref{fig:black-scholes-RRs}), in the SIR model (\subref{fig:SIR-RRs}) and in the richer economic model (\subref{fig:stochasticDomination}).}

  \label{fig:cdc_vs_collective_drawdowns}
\end{figure}

A fan diagram of the outcome of the dynamic-accrual designs we chose and
the comparable collective drawdown designs are shown in Figure 
\ref{fig:stochasticDomination}. We plot all the deciles from 10\% to 90\% in each figure. 
The choice of loss function is ultimately subjective, but we 
believe that this approach should ensure our choice of loss function leads
to risk levels that would be found acceptable within the pension industry.

As can be seen from Figure \ref{fig:cdc_vs_collective_drawdowns}, collective drawdown outperforms the dynamic-accrual CDC in all economic models. In particular, in the richer economic model,
the collective-drawdown approach can (at least to the accuracy shown in the diagram) stochastically dominate the dynamic-accrual design (see panel \subref{fig:stochasticDomination}). This shows not only the success of our approach, but also that when optimising over a more complex model, there is potentially more to gain.

A key advantage of the machine-learning approach is that
essentially the same code can be used for different economic models.
We have compared the strategy found using machine learning in the Black--Scholes model with the optimal strategy computed using a classical approach which can be proved to converge. The discrepancy between the two approaches is small: less than 1\% when measured using certainty equivalents. This gives some confidence that our approach is able to find good quality strategies.

To find the optimal strategy, each neural network was trained ten times for 24 hours on a high performance computing system and we selected the best performing strategy. The additional improvements found through this process explain the improved values given in this paper compared to our report \cite{armstrongPike}.

With minimal changes, the same code used to optimize a collective-drawdown design can be used to find the optimal DC strategy followed by annuitisation, the optimal DC strategy followed by flex-and-fix and the optimal DC strategy followed by drawdown.

\medskip

For shared-indexation designs, there is limited flexibility for scheme designer to optimize. Consumption in retirement is fixed by the rules on nominal benefits. The pooling of members' assets in shared-indexation designs makes it challenging to know how one would go about dynamically changing investment strategy, in particular if one wishes to keep cross-subsidies to a minimum one needs to change the prices charged as one changes strategy.

For this reason, we have limited our search for optimal CDC strategies to (i) varying the start and end date of the lifestyle-strategy used to determine the investment strategy and (ii) for flat-accrual schemes adjusting the target level of indexation. The strategies we considered assume a member invests at 100\% in the risky asset until a given age, then tapers linearly to 0\% in risky assets at some subsequent age. One could, of course, design more complex lifestyle curves but
when the fund is in a long-term steady state, the details of this strategy would be unimportant: what matters is the overall proportion invested in risky assets by the scheme.  With this in mind, we further reduced the dimension of our search space by requiring that the start age of tapering was $55+x$ and the end age for tapering was $75+2x$ for a given choice of $x$. For flat-accrual schemes we evaluated the effect of changing the target level of indexation by considered the target value $\CPI$, $\CPI+0.5\%$ and $\CPI+1.0\%$.


To obtain a loss function for the scheme as a whole, we took the greater of the loss function for generations 40 and generation 95. The fund is in an approximately steady state between these generations, so our intention is to identify the worst case within this steady-state. Earlier generations are not in the fund for their working full career, so it is difficult to include them in a comparison. Later generations experience some effects from the fund closing. We then sought the lifestyle-strategy and target indexation which minimized this loss.

We accept that it is possible to imagine a range of more complex modifications to the shared-indexation design one could try, but we are not aware of any concrete proposals. Moreover, the intention of the shared-indexation design is that it is simple to explain, so the more one complicates the design the less convincing it is to require imposing the shared-indexation constraint.

\section{Results}\label{sec:results}

The results of our simulations are summarised in \Cref{table:mainResults}. For each scheme design we have computed the {\em certainty-equivalent replacement ratio} for every economic model. This
is defined to be the constant value $\gamma \in {\R}$ solving
\[
U(\overline{\gamma})=U(c)
\]
where $\overline{\gamma}_t:=\gamma$ for all $t$ and $c$ is the random-consumption process of the optimal design.

\begin{sidewaystable}[h!tbp]
\begin{center}
\makebox[\textwidth][c]{%
\begin{tabular}{
p{20mm}
p{38mm}
>{\centering\arraybackslash}p{22mm}
>{\centering\arraybackslash}p{28mm}
>{\centering\arraybackslash}p{26mm}
} 
\toprule
{\bf Group} & {\bf Design} & 
\multicolumn{3}{c}{\bf Certainty-equivalent replacement ratio} \\ 
\cmidrule(lr){3-5}

& & 
\makecell{Black--Scholes}
& \makecell{SIR Model}
& \makecell{Richer \\ Economic Model} \\ 

\midrule

\multirow{2}{20mm}{Individual} 
& Optimal DC + Annuity & 39.9\% & 52.6\% & 55.8\% \\
& Optimal Flex and Fix & 43.9\% & 63.5\% & 67.4\%\\ 
\midrule

\multirow{3}{20mm}{Shared indexation} 
& Flat accrual & 40\% & 40\% & 45\% \\
& Dynamic accrual & 35\% & 37.5\% & 45\% \\
\midrule

N/A & Collective drawdown & 45.7\% & 71.8\% & 86.6\% \\ 

\bottomrule
\end{tabular}
}
\end{center}
\caption{Optimal certainty-equivalent replacement ratio for different scheme
designs. We present the neural-network optimised strategies to one decimal place and the CDC simulations to the nearest $2.5\%$. This reflects the approximate accuracy of the calculations: CDC simulations are expensive to perform limiting the accuracy of our Monte Carlo simulations.}
\label{table:mainResults}
\end{sidewaystable}

It is unsurprising that the collective drawdown design performs the best.
It is trivial to prove that collective drawdown must outperform the flex-and-fix strategy which must, in turn, outperform annuitisation. We cannot prove
that collective drawdown must outperform shared-indexation in the richer economic model since this is an incomplete
market model, but we can prove the analogous results in complete markets (e.g. the Black-Scholes model or a continuous time version of the SIR model).
Since none of the other shared-indexation designs contain any features designed to exploit market
incompleteness, it would be remarkable if they outperformed collective drawdown in the richer economic model.

Another readily explained feature of our results is that the performance gap increases with the richness of the economic model. This is because the collective drawdown design is able to optimize against any features of the economic model but other designs are more rigid. The shared-indexation designs are the most rigid and so benefit the least.

There is a risk of over-fitting: if the economic model contains features that are not present in reality one may over-estimate the potential for optimization. We have attempted to mitigate the risk of over-fitting by limiting the investment strategies available to just an equity index and a bond index. In our view, because of its assumption of constant interest rates, using the Black--Scholes model will under-estimate the difference between designs. The rich economic model on the other hand is something of a black box, and so there is a risk of it over-estimating the differences. The SIR model is intended to err slightly on the side of under-estimating the difference between designs while maintaining parsimony.


It is not easy to predict the relative rankings of flex-and-fix and the shared-indexation designs. flex-and-fix allows greater flexibility in the strategy, but shared-indexation designs allow longevity risk to be fully hedged when investing in risky assets. It is hard to guess a-priori which is more important and one might well see different behaviour if one varied the choice of risk preferences.

Similarly, it is hard to guess which will perform better, flat-accrual designs or dynamic-accrual designs. Flat-accrual can suffer from drag effects which lead to inefficiency, but dynamic-accrual is less flexible as it does not allow the long-term level of indexation and the risk of the scheme to be adjusted independently. For our
choice of risk-preferences the two designs are essentially tied.

One caveat on our results is that we have not included systematic mortality risk in our modelling.
In the presence of systematic mortality risk one must pay an additional premium to purchase an annuity, so the designs we consider that use annuitisation will then perform less well. Systematic longevity risk will also affect collective designs, but the evidence from \cite{armstrong_dalby} suggests that this effect will not be as significant as for designs involving annuities. This is because it is possible to adapt one's consumption strategy to benefit from new information about mortality in these designs, whereas an annuity pre-commits to fixed payments. The later one annuitises the less systematic longevity risk that will be present.

A second caveat on our results is that we cannot be sure how close the strategies we have found
are to the true optimum in the richer economic model. In the Black--Scholes model, it is feasible to calculate the true optimal strategy using a provably convergent numerical method and we have used this to test that our machine-learning approach gives close to optimal results for decumulation-only investments. For more complex models, for example our richer economic model, our forward-approach may be too simplistic and one may find better results using alternative state-of-the-art algorithms. Ultimately, as the true answer is unknown, all one can do is experiment with a selection of algorithms and select the one that performs the best. Thus one should think of our results in the richer economic model as giving lower bounds on the performance of strategies subject to the limitations of the algorithms we have used and the processing power we have available.

A third caveat is that we have used some approximations in our modelling of term structures. We allow each design to invest in a bond index that matches the relevant liability, namely an annuity (case (i)) or the liabilities of the fund (case (ii)). As a result, we had to use an approximation to the term-structure for shared-indexation designs in the SIR model and to all designs in the richer economic model. We do not expect this to have had a material effect on results, but simply wish to remind the reader where approximations have been made.

\subsection{Optimal investment strategies}

\subsubsection{Collective Drawdown}

The investment strategy for collective drawdown, across all of the economic models, is to take a highly leveraged position when one is young. The optimal investment and consumption strategies are shown in Figures
\ref{fig:collectiveDrawdownStrategy_blackScholes}, \ref{fig:collectiveDrawdownStrategy_blackscholesSIR} and \ref{fig:collectiveDrawdownStrategy_ppiEsg} in the appendix.

This leverage decreases rapidly with age. Although an individual's investment strategy is leveraged, the fund as a whole in a given year may not be. Indeed, as member's total wealth increases during their lifetime, the overall position of the fund will not typically be highly leveraged.

To test the importance of the very high leverage seen at the beginning of the investor's working life, we evaluated the same investment and consumption strategy but applying a cap on leverage. When leverage was capped at 300\% we found that the certainty equivalent reduced by approximately 0.59\%, 0.64\% and 1.14\% in the economic models respectively and when it was capped at 200\% it reduced by approximately 1.79\%, 2.73\% and 3.90\%. We conclude that while some leverage is an important part of the strategy, such extreme leverage is not important.

When compared to the optimal strategies found
in the Black-Scholes model, we see greater variability
in the optimal proportions to invest
and consume each year when they are computed
in this richer economic model. This is unsurprising
as the state of the richer economic model contains additional information that can be exploited to
find superior strategies. For example, as interest
rates are mean reverting in the richer model one would
expect the optimal investment proportions to vary
as interest rates change as this will change the
market price of risk. A similar pattern is seen in all the scheme comparisons as we vary the economic model.

\subsubsection{Annuity-based strategies}

High leverage early in the life-course is seen in the optimal strategy for
DC followed by full annuitisation (Figures \ref{fig:annuityStrategy_BlackScholes}, \ref{fig:annuityStrategy_SIR} \& \ref{fig:annuityStrategy_ppiEsg}.)

The optimal flex-and-fix strategy (Figures \ref{fig:flexThenFix_BlackScholes}, \ref{fig:flexThenFix_SIR} \& \ref{fig:flexThenFix_ppiEsg}) is to annuitise the majority of one's wealth at retirement and then take a leveraged position with
the remaining assets. This achieves a net equity exposure similar to that of a collective drawdown design.

In a collective design, highly leveraged positions can
be synthesised within the fund itself. Achieving similarly leveraged
positions in a DC design will be more challenging and is likely
to increase the level of transaction costs. Moreover, the explicit use of leverage may make it difficult to achieve regulatory approval for any design.

\subsubsection{Shared-indexation designs}

We have chosen our gain function so that it gives an optimal collective
drawdown strategy whose results are comparable to a dynamic-accrual strategy which invests in an asset mix similar to that suggested by our industry consultation.
As one can see from Figure \ref{fig:stochasticDomination} the collective drawdown design has a similar upside to a dynamic-accrual design early in retirement, but a somewhat better downside. As a result, one might expect that optimizing the shared-indexation designs
will result in lower-risk strategies being adopted to avoid this unwanted downside. This is indeed the case: the optimal flat-accrual design invests a median amount
of 46\% in the risky asset in its steady state;
the optimal dynamic-accrual design invests a median amount of 58\% in the risky
asset in its steady state.

The replacement ratios in retirement for the optimal flat and dynamic-accrual designs are plotted in Figure \ref{fig:sharedIndexationFigure}. Both designs
result in median incomes which increase in real-terms in retirement. The optimization procedure for the flat-accrual design has the flexibility to choose a target level of indexation equal to $\CPI+1\%$.

The dynamic-accrual design cannot independently vary the asset mix and the long-term level of indexation and
so it must strike a balance between the optimal risk level and optimal indexation.
This lack of flexibility in the dynamic-accrual design means it is only able to perform comparably to the flat-accrual design from the point of view of our gain function, despite the flat-accrual design having drag effects and greater disparities between the best-performing and worst-performing generations.

\section{Conclusion}

If one assumes that a scheme member has no interest in leaving a bequest,
our results show that 
collective drawdown performs better than the other designs when assessed purely in terms
of member utility. It does so across three separate economic models.

The lack of a bequest may well be a concern for some investors.
However, this is equally true of any other design which
features longevity pooling and so should, in principle, apply equally to shared-indexation designs.
Nevertheless, the collective drawdown approach makes the longevity pooling much more explicit
and this may be unattractive to members. The importance (or otherwise) of this issue will depend upon
how the design is presented to members.

Another consideration that is important in designing a scheme is whether members understand how a scheme operates.
We believe that the collective drawdown design is much simpler to understand than shared-indexation designs.

Within the UK market it is important that members should be able to opt out of any scheme, at least before retirement. This is facilitated in a collective drawdown design (and in DC designs) by the use of a separate account for each member. This makes transfer values trivial to compute and eases regulation of the scheme.

We do not wish to propose a detailed design for a collective drawdown product in this paper, that is a commercial matter, but let us mention some of the issues to consider. A collective drawdown approach could be used for a whole-life product or as a decumulation-only product. In the UK, this choice would determine whether the product was regulated by the Pensions Regulator or the FCA.
One should also consider how much choice members should have over investment and consumption strategies. Members may prefer to choose a strategy from a small set of options, even though this limits choice. It may also be advisable to prevent members selecting inappropriate strategies where they will run out of money.

In summary, collective drawdown outperforms shared-indexation designs and enjoys a number of
additional practical advantages over these designs. Key outstanding questions are how collective
drawdown products should be designed and presented to ensure that they are
attractive to consumers and will meet their needs.

\section*{Acknowledgments}

JA and JD gratefully acknowledge funding from the Nuffield Foundation (grant FR-000024058). RH is funded by EPSRC DTP (grant ref: EPSRCDTP2401).

\bibliographystyle{plain}
\bibliography{bib}

\appendix

\section{Computation of the loss function}
\label{sec:lossComputation}

To compute the value of the gain function, we assume that consumption and individual longevity risk are independent. We also assume there is no systematic longevity risk and that the unconditional probability an individual dies in a given year $s$ is given by $\overline{p}_s$.
\begin{equation}
    U = -\mathbb{E}_{\mathrm{Invest}} \left[ \sum_{s=t_{\RA}}^T \overline{p}_s\, \exp\left(-\alpha\sum_{t=t_{\RA}}^{s}  u(C_t) \delta t\right) \right].
\end{equation}
In this formula, ${\mathbb E}_\mathrm{Invest}$ denotes the expectation across investment scenarios
and so excludes the mortality component of our probability model, which is accounted for by the term ${\overline{p}}_s$. $T$ is the maximum time of death, which is finite for the mortality model we are using. If we generate $N$ investment scenarios and label the replacement ratio in each case $C^{(j)}$ with $1\leq j \leq N$, we may estimate the gain function using the following expression

\begin{equation}
    \hat{U}:=-\frac{1}{N} \sum_{j=1}^N \left[ \sum_{s=t_{\RA}}^T \overline{p}_s\, \exp\left(-\alpha\sum_{t=t_{\RA}}^{s}  u(C_t^{(j)}) \delta t \right) \right].
    \label{eq:utility}
\end{equation}

Since our EKM gain function is inhomogeneous, maximising it
using traditional PDE techniques could be computationally demanding. This is because
we do not benefit from the dimension reduction that occurs from using homogeneous preferences. 
As such, we have chosen to use machine learning, specifically a recurrent neural network, to find the optimal strategy that maximises this gain function. In this setting, the problem can be formulated straightforwardly as requiring the network to learn the controls that maximise the gain function $\hat{U}$. We give technical details on the training and use of our network in Appendix \ref{appendix:neuralNetworkDesign}.

We take $L=\log(-\hat{U})$ to be the loss function
for our machine learning model, so that minimising $L$ is equivalent to maximising $\hat{U}$. The $\log$ transform is chosen to avoid excessively large loss values. First note that we may write
\begin{align*}
    -\hat{U} &= \frac{1}{N}\sum^N_{j=1}\sum_{s=t_{\RA}}^T \exp\left(\log(\overline{p}_s)-\alpha\sum^s_{t=t_{\RA}}u(C_{t}^{(j)}) \delta t\right)\\
    & = \frac{1}{N}\sum^N_{j=1}\exp\left(\log\left(\sum_{s=t_{\RA}}^T \exp\left(\log(\overline{p}_s)-\alpha \sum^s_{t=t_{\RA}}u(C_{t}^{(j)}) \delta t\right)\right)\right).
\end{align*}
Hence, our loss function is given by
\begin{multline}\label{eq:loss}
    L = \log\left(\sum^N_{j=1}\exp\left(\log\left(\sum_{s=t_{\RA}}^T \exp\left(\log(\overline{p}_s)-\alpha \sum^s_{t=t_{\RA}}u(C_{t}^{(j)}) \delta t\right)\right)\right)\right)\\
    - \log(N).
\end{multline}
We then evaluate this expression using a $\logsumexp$ function to avoid excessive rounding errors.

\section{Continuous time extension of the richer economic model}
\label{sec:ctsTimeESG}

To find an appropriate continuous-time version of our discrete-time model we first introduce the following definitions.

Let ${\cal A}_n$ denote the space of $n \times n$ matrices, $A$,  with $\ker A \cap\im A =\{0\}$. Let $\hat{A}^{-1}$ denote the transformation equal to $\left(\hat{A}_{|\im \hat{A}}\right)^{-1}$ on $\im \hat{A}$ and equal to $0$ on $\ker \hat{A} $.
Let $\pi^k_{\hat{A}}$ denote the projection onto the kernel of $\hat{A}$ and $\pi^i_{\hat{A}}$ the projection onto the image.

The next proposition allows us to compute a continuous time
version of the process $\mathbf{x}_t$. 

\begin{proposition}
\label{prop:ctsTimeVersion}
Let $\mathbf{x}$ be a process satisfying equation \eqref{eq:discreteVAR}.
Suppose that $A \in {\cal A}_n$.
Suppose that the $n\times n$ matrix $\hat{A}$ and the vector $\hat{\mathbf{b}} \in \R^n$
satisfy
\begin{subequations}
\begin{align}
    \hat{A} &=     \log(A+\mathbf{I})\in\mathbb{R}^{n\times n},\\ 
(\pi^k_A + A)\hat{\mathbf{b}}
&=
(\pi^k_A + \hat{A})\mathbf{b} \label{eq:bHat}.
\end{align}
\end{subequations}
Define a rank-4 tensor $\Gamma$ by
\begin{equation}
    \Gamma_{iajb} = \int_0^1 \exp(\hat{A}(1-u)_{ia})\exp(\hat{A}(1-u)_{jb})du.
\label{eq:gammaTensor}
\end{equation}
Suppose that $\Gamma$ is such that the equation 
\begin{equation}
\sum_{a,b} \Gamma_{iajb}\hat{\Sigma}_{ab} = \sum_a L_{ia} L_{ja}.
\label{eq:linear2}
\end{equation}
has a unique symmetric solution $\Sigma_{ab}$ which is positive semi-definite. We may then define $\hat{L}$ by solving
$\hat{L} \hat{L}^T = \hat{\Sigma}$.

Under these circumstances, if $\hat{\mathbf{x}}_t$ has dynamics given by
\begin{equation}
    \mathrm{d} \hat{\mathbf{x}}_t = (\hat{A}\hat{\mathbf{x}}_t + \hat{\mathbf{b}})\mathrm{d}t + \hat{L}\mathrm{d}\textbf{W}_t \label{eq:esg-sde}
\end{equation}
then $\hat{\mathbf{x}}$ at integer times it will be identically distributed
to $\mathbf{x}$.
\end{proposition}

We prove this Proposition in \Cref{sec:continuous-ESG}.
One can compute the
tensor $\Gamma$ using numerical
integration, and we used Gaussian quadrature with 20 integration points. The equations \eqref{eq:bHat} and \eqref{eq:linear2} are simply linear equations so it is easy to find solutions if they exist.

Having identified the SDE \eqref{eq:esg-sde}, we can then compute the equations for the log price of assets using It\^o's Lemma. If we have assets of price $P_i$, for $i=1,\ldots,n$, that satisfy
\begin{equation}
\mathrm{d}\log(P_i)=\mathbf{\mu}_i (t, P_1, \ldots, P_n)\mathrm{d}t + \sum^d_{j=1}\sigma_{i,j}(t, P_1, \ldots, P_n)\mathrm{d}\mathbf{W}^j_t,
\end{equation}
where $\mu\in\mathbb{R}^n$, $\sigma\in\mathbb{R}^{n,d}$, $\mathbf{W}\in\mathbb{R}^d$ is a $d$-dimensional Brownian motion, and for investment proportions of one's wealth $\pi_i$ in asset $i$, the realised drift and volatility can be substituted into the log wealth equation
\begin{equation}
    \mathrm{d}\log (w_{t+\delta t}) = \sum^n_{i=1} \pi_i \left(\mu_i + \frac{1}{2} \sigma_{i,\alpha}\sigma_{i,\alpha}\right)\mathrm{d}t - \frac{1}{2}\sum^n_{i,j}\pi_i \pi_j \sigma_{i,\alpha} \sigma_{j,\alpha} \mathrm{d}t + \pi_i \sigma_{i,\alpha} \mathrm{d}\mathbf{W}^\alpha_t 
\label{eq:logWealth}
\end{equation}
to simulate investment returns. Here we use the Einstein summation convention for ease of notation.

In practice a collective scheme would observe the desired net position of all members and follow an investment strategy that reflects these preferred investment weights as closely as feasible without incurring excessive transaction costs. The value of each members' fund at the end of the period could then be computed as a proportion of the total final position of the fund, with the proportion received by each member determined by the ratio of the values of $w_{t+\delta t}$ for each member, as computed using equation \ref{eq:logWealth}. In this way a collective scheme can allow some members to take a large leveraged position.

This is the simplest mechanism of allowing for leveraged investments within the fund. For the purposes of this paper, we assume that members are willing to tolerate some mispricing of leveraged positions that will inevitably occur if they are not calibrated to market option prices. One can view it as part of the collective agreement that members mutually agree to allow one another to take leveraged positions even if they are slightly mispriced because of the net benefit everyone receives by allowing leverage. Ideally, one might try to price leveraged positions as accurately as possible by calibrating to market option prices, but we will not attempt to model this.


\section{Proof of Proposition \ref{prop:ctsTimeVersion}}
\label{sec:continuous-ESG}

\begin{lemma}
If $\hat{\mathbf{x}}_t$ has dynamics given by equation \eqref{eq:esg-sde}
and $\hat{A} \in {\cal A}_n$
then
\begin{equation}
\hat{\bf{x}}_t
= \exp(\hat{A}t)\left[
\hat{\bf{x}}_0 + \hat{A}^{-1} \hat{\bf{b}} \right] + \int_0^t
 \exp(\hat{A}(t-s)) L \, d \mathbf{W}_s 
 - \hat{A}^{-1} \hat{\bf{b}} + \pi^{k}_{\hat{A}} \hat{\bf{b}} t.
\label{eq:sdeSolution}
\end{equation}
\end{lemma}
\begin{proof}
Define 
\begin{equation}
{\bf z}_t:= \exp( -\hat{A} t) (\hat{\bf{x}}_t + \hat{A}^{-1} \hat{\bf{b}}) - \pi^{k}_{\hat{A}} \hat{\bf{b}} t.
\label{eq:defz}
\end{equation}
We compute
\begin{align}
d {\mathbf z}_t
 &= -\hat{A} \exp(-\hat{A} t)(\hat{\mathbf x}_t + \hat{A}^{-1} \hat{\mathbf b} ) dt
 - \pi^{k}_{\hat A} \hat{\mathbf b} \, dt + \exp(-\hat{A}t) \, d \hat{\mathbf x}_t \nonumber \\ 
&= \left\{ \exp(-\hat{A} t)
\left[ -\hat{A} ({\mathbf x}_t + \hat{A}^{-1} \hat{\mathbf b} )
+ \hat{A} {\mathbf{x}}_t + \hat{\mathbf b} 
\right]  - \pi^{k}_{\hat A} \hat{\mathbf b} \right\} dt 
+
\exp(-\hat{A} t) \hat{L} \, d {\bf W}_t
\label{eq:beforeSimplification}
\end{align}
using the fact that $\hat{A}$ commutes with $\exp( -\hat{A} t)$ 
and the equation \eqref{eq:esg-sde}.
Since $\exp(-\hat{A}t)$ is equal to the identity on the kernel of $\hat{A}$ we have
\begin{equation}
\exp(-\hat{A}t) \pi^k_{\hat{A}} \hat{\bf b} = \pi^k_{\hat{A}} \hat{\bf b}.
\label{eq:expOnKernel}
\end{equation}
Our definition of $\hat{A}^{-1}$ ensures that
\[
\hat{A} \hat{A}^{-1} \hat{\mathbf{b}}
 = \pi^i_{\hat{A}} \hat{\mathbf{b}}.
\]
We therefore find that
equation \eqref{eq:beforeSimplification}
simplifies to
\begin{align*}
d {\bf z}_t
 &=  \exp(-\hat{A} t)
\left[ \pi^{i}_{\hat{A}}\hat{\mathbf{b}}
+ \pi^{k}_{\hat{A}} \hat{\mathbf{b}}
- \hat{\mathbf{b}} \right] dt
+
\exp(-\hat{A} t) \hat{L} \, d {\bf W}_t \\ 
&= \exp(-\hat{A} t) \hat{L} \, d {\bf W}_t. 
\end{align*}
Thus
\[
\mathbf z_t
 = \mathbf z_0 + \int_0^t \exp(-\hat{A}s ) \hat{L} \, d \mathbf{W}_s.
\]
Substituting this into equation
\eqref{eq:defz}
gives equation \eqref{eq:sdeSolution}.
\end{proof}

\begin{lemma}
\label{lemma:differenceEquation}
If $\hat{\mathbf{x}}_t$ has dynamics given by equation \eqref{eq:esg-sde}
and $\hat{A} \in {\cal A}_n$
then $\hat{\mathbf{x}}_t$
satisfies the difference equation
\begin{equation}
\hat{\mathbf{x}}_{t+1}-\hat{\mathbf{x}}_t
= (\exp(\hat{A})-1)(\hat{\mathbf{x}}_t + \hat{A}^{-1} \hat{\mathbf{b}})
+ \pi^{k}_{\hat{A}} \hat{{\mathbf{b}}}
+ \int_0^{1} \exp(\hat{A}(1- s)) L \, d\mathbf{W}_{t+s}.
\end{equation}
\end{lemma}
\begin{proof}
Using equation \eqref{eq:sdeSolution} we have
\begin{align*}
\hat{\mathbf{x}}_{t+1} - \hat{\mathbf{x}}_t
&= \exp(\hat{A}(t+1))\left[
\hat{\bf{x}}_0 + \hat{A}^{-1} \hat{\bf{b}} + \int_0^{t+1}
 \exp(-\hat{A}s) L \, d \mathbf{W}_s \right] \\
&\quad
 - \hat{A}^{-1} \hat{\bf{b}} + \pi^{k}_{\hat{A}} \hat{\bf{b}} (t+1) - \hat{\mathbf{x}}_t \\
&= \exp(\hat{A}) \exp(\hat{A} t)\left[
\hat{\bf{x}}_0 + \hat{A}^{-1} \hat{\bf{b}} + \int_0^{t}
 \exp(-\hat{A}s) L \, d \mathbf{W}_s
 \right] \\
&\quad
 - \hat{A}^{-1} \hat{\bf{b}} + \pi^{k}_{\hat{A}} \hat{\bf{b}} (t+1) - \hat{\mathbf{x}}_t 
 + \exp(\hat{A}(t+1)) \int_t^{t+1}
 \exp(-\hat{A}s) L \, d \mathbf{W}_s.
\label{eq:sdeSolution}
\end{align*}
Using equation \eqref{eq:sdeSolution} a second
time we obtain
\begin{align*}
\hat{\mathbf{x}}_{t+1} - \hat{\mathbf{x}}_t
&= \exp(\hat{A}) \left[
\hat{\bf{x}}_t + \hat{A}^{-1} \hat{\bf{b}} - \pi^k_{\hat{A}} \hat{\mathbf{b}} t\right] \\
&\quad
 - \hat{A}^{-1} \hat{\bf{b}} + \pi^{k}_{\hat{A}} \hat{\bf{b}} (t+1) - \hat{\mathbf{x}}_t 
 + \exp(\hat{A}(t+1)) \int_t^{t+1}
 \exp(-\hat{A}s) L \, d \mathbf{W}_s\\ 
&= (\exp(\hat{A})-1) (
\hat{\bf{x}}_t + \hat{A}^{-1} \hat{\bf{b}}) + \pi^k_{\hat{A}} \hat{\mathbf{b}}
 + \int_0^{1}
 \exp(\hat{A}(1-s)) \hat{L} \, d \mathbf{W}_s.
\end{align*}
We have used equation
\eqref{eq:expOnKernel}
when simplifying the final line.
\end{proof}
By the It\^o isometry, the term 
\[
 \exp(\hat{A}) \int_0^{1}
 \exp(-\hat{A}s) \hat{L} \, d \mathbf{W}_s
\]
has a multivariate normal distribution with mean
$0$ and covariance matrix
\begin{equation}
  \int_0^{1}
 \exp(\hat{A}(1-s)) \hat{L} \hat{L}^\top \exp(\hat{A}(1- s))^\top \, dt.
\label{eq:covarianceMatrix}
\end{equation}
This motivates us to define the tensor $\Gamma$ using
equation \eqref{eq:gammaTensor}.

The equation \eqref{eq:linear2}
can be thought of as a linear
equation on ${\mathbf \R^d \times \mathbf \R^d}$ so it is straightforward to solve for
$\hat{\Sigma}$, if a solution exists. One can now find $\hat{L}$ satisfying $\hat{L} \hat{L}^T = \hat{\Sigma}$ by Cholesky decomposition. If we compute $\hat{L}$ in this way then the covariance matrix
\eqref{eq:covarianceMatrix}
will equal $L L^\top$. Let us assume this holds for the remainder of this section.

Thus Lemma \ref{lemma:differenceEquation}
tells us that $\mathbf{x}_t$ 
at integer times will
be identically distributed 
to $\mathbf{x}_t$ so long as
\begin{equation}
\exp(\hat{A})-1 = A,
\label{eq:expEquation}
\end{equation}
and
\begin{equation}
A \hat{A}^{-1} \hat{\mathbf{b}}
+ \pi^{k}_{\hat{A}} \hat{{\mathbf{b}}}
 = {\mathbf{b}}.
 \label{eq:bEquation}
\end{equation}
If we set $\hat{A} = \log(A + 1 )$
equation \eqref{eq:expEquation} is assured. This also ensures that the kernel and image of $\hat{A}$ is equal to the kernel and image of $A$ and so that $\hat{A} \in {\cal A}_n$.
This also allows us to rewrite \eqref{eq:bEquation} as
\[
(A + \pi^k_A) (\hat{A}^{-1} 
+ \pi^{k}_{\hat{A}}) \hat{{\mathbf{b}}}
 = \mathbf{b}.
\]
Using the commutativity of all the operators this may be written as
\[
(\hat{A}^{-1} 
+ \pi^{k}_{\hat{A}})(A + \pi^k_A) \hat{{\mathbf{b}}}
 = \mathbf{b},
\]
and then further rearranged to give
equation \eqref{eq:bHat}.
This completes the proof of Proposition \ref{prop:ctsTimeVersion}.

\section{Network architecture and implementation}
\label{appendix:neuralNetworkDesign}

Our machine learning code base is built in Python using the Tensorflow package.
We use a recurrent neural network consisting of six layers: 
\begin{itemize}
    \item The first (input) layer has a set number of nodes that depends on the economic model. The nodes effectively each represent a different piece of data. Specifically, these include time, mortality and all the economic data we require e.g., wage growth, inflation, the stock price etc.

    \item The second layer is a dense layer with 80 nodes and a ReLU activation function.

    \item The third layer is a gated recurrent unit (GRU) with 25 nodes. This is the recurrent layer in our network. The activation function is the hyperbolic tangent function and the recurrent activation function is sigmoid. This layer allows us to deal with the time dependent nature of our data. The GRU is set up so as to return an output at each time point, hence the final layer returns predictions at each time point.
    
    \item The fourth and fifth layers are identical to the second layer.

    \item For the final layer, if we have $n$ assets to invest in, then we have $$(n-1) +1$$ nodes. Each node represents a control we have over the problem. The $n-1$ controls represent the investment proportions to choose. Since proportions must sum to 1, one proportion is automatically determined as 1 minus the total investment proportion. The remaining control is the proportion of wealth to consume. Because of the way the GRU is used, we therefore obtain investment and consumption decisions in each year of our simulation. Consumption before retirement is ignored by our loss function.

\end{itemize}
\noindent
We used the Adam optimizer with an initial learning rate of 0.001. Training was carried out for 24 hours, each epoch consisting of 65536 scenarios with a batch size of 4096. We simply let the training run and then take the weights of the model from the epoch with the least loss. A validation set of 10,240 separately generated scenarios was evaluated at the end of each epoch. This gave the strategies in \Cref{sec:results}.

\section{Optimal investment strategies}

\pagebreak
\subsection{Collective Drawdown}

\begin{figure}[h!tbp]
    \centering
    \begin{minipage}{0.625\textwidth}
    \centering
        \includegraphics[width=1.0\textwidth]{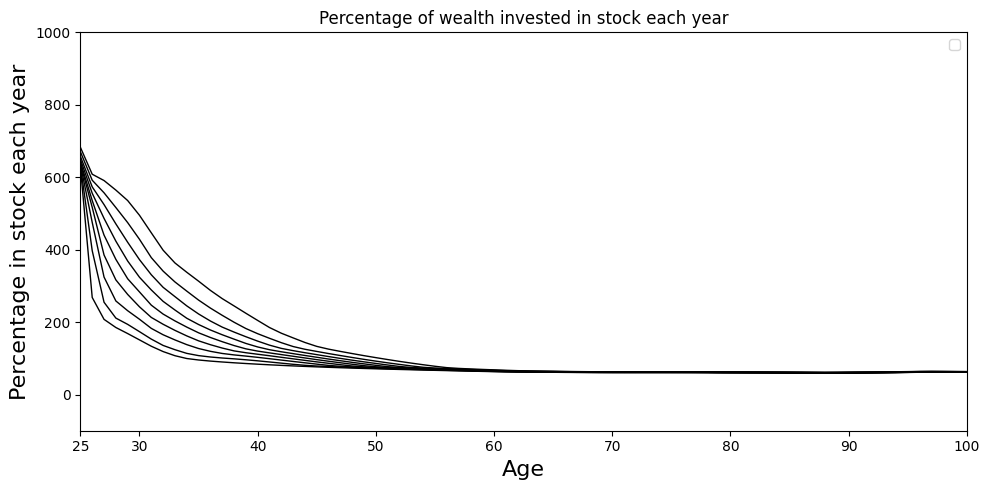}
    \end{minipage}
    \begin{minipage}{0.365\textwidth}
    \centering
        \includegraphics[width=1.0\textwidth]{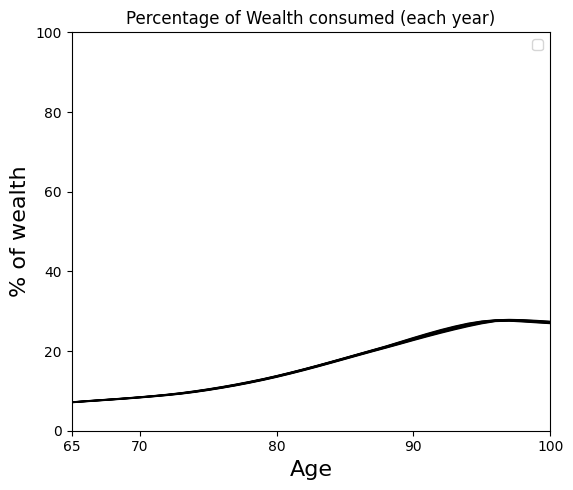}
    \end{minipage}
        \caption{Fan diagrams for the optimal collective drawdown strategy in the Black--Scholes model, showing deciles of: the percentage of wealth spent on the stock; the percentage of wealth consumed.}
    \label{fig:collectiveDrawdownStrategy_blackScholes}
\end{figure}
\begin{figure}[h!tbp]
    \centering
    \begin{minipage}{0.625\textwidth}
    \centering
        \includegraphics[width=1.0\textwidth]{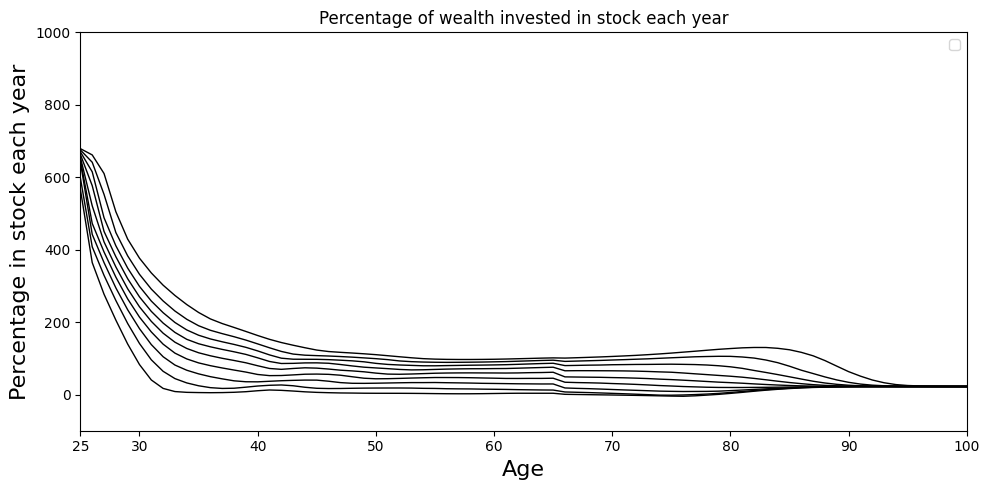}
    \end{minipage}
    \begin{minipage}{0.365\textwidth}
    \centering
        \includegraphics[width=1.0\textwidth]{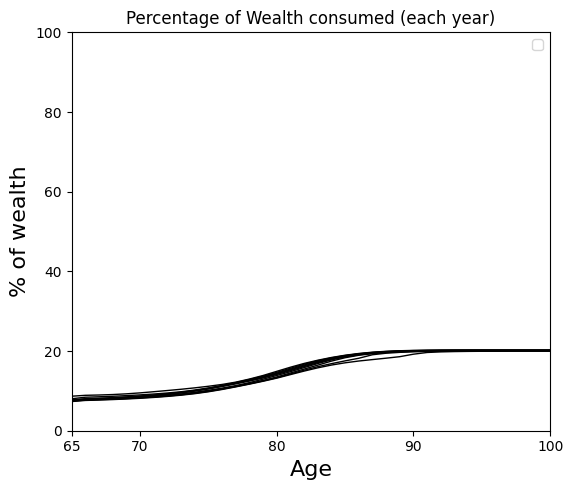}
    \end{minipage}
        \caption{Fan diagrams for the optimal collective drawdown strategy in the SIR model, showing deciles of: the percentage of wealth spent on the stock; the percentage of wealth consumed.}
    \label{fig:collectiveDrawdownStrategy_blackscholesSIR}
\end{figure}
\begin{figure}[h!tbp]
    \centering
    \begin{minipage}{0.625\textwidth}
    \centering
        \includegraphics[width=1.0\textwidth]{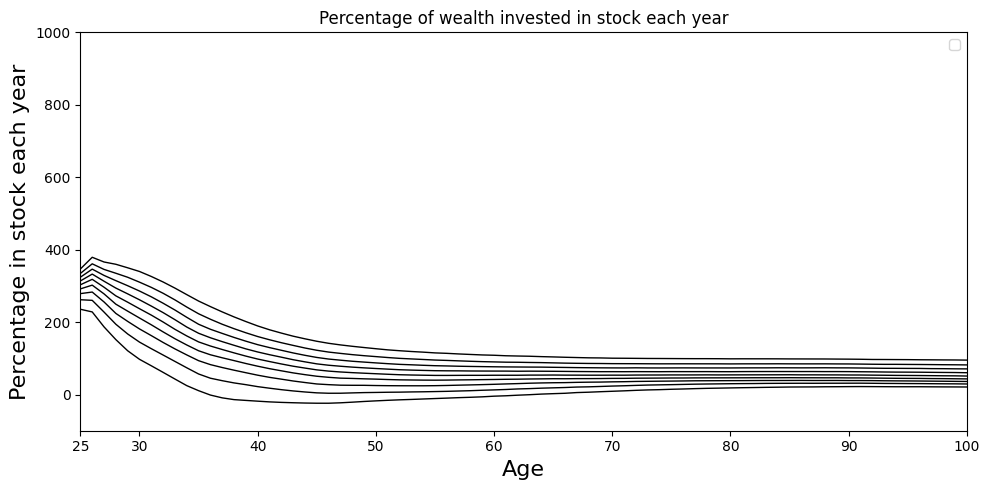}
    \end{minipage}
    \begin{minipage}{0.365\textwidth}
    \centering
        \includegraphics[width=1.0\textwidth]{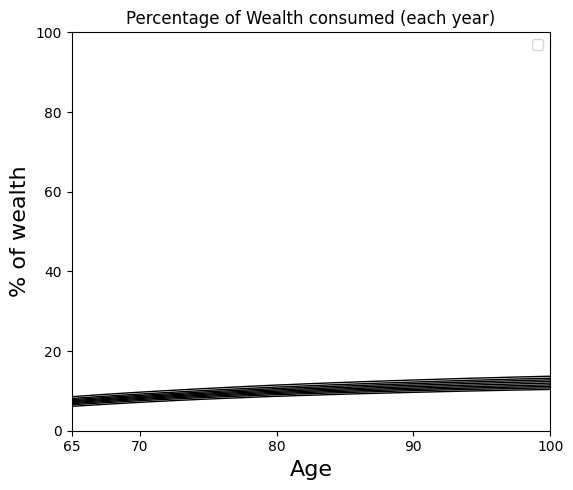}
    \end{minipage}
        \caption{Fan diagrams for the optimal collective drawdown strategy in the richer economic model, showing deciles of: the percentage of wealth spent on the stock; the percentage of wealth consumed.}
    \label{fig:collectiveDrawdownStrategy_ppiEsg}
\end{figure}

\pagebreak
\subsection{Full annuitisation}

\begin{figure}[h!tbp]
    \centering
    \begin{minipage}{0.625\textwidth}
    \centering
        \includegraphics[width=1.0\textwidth]{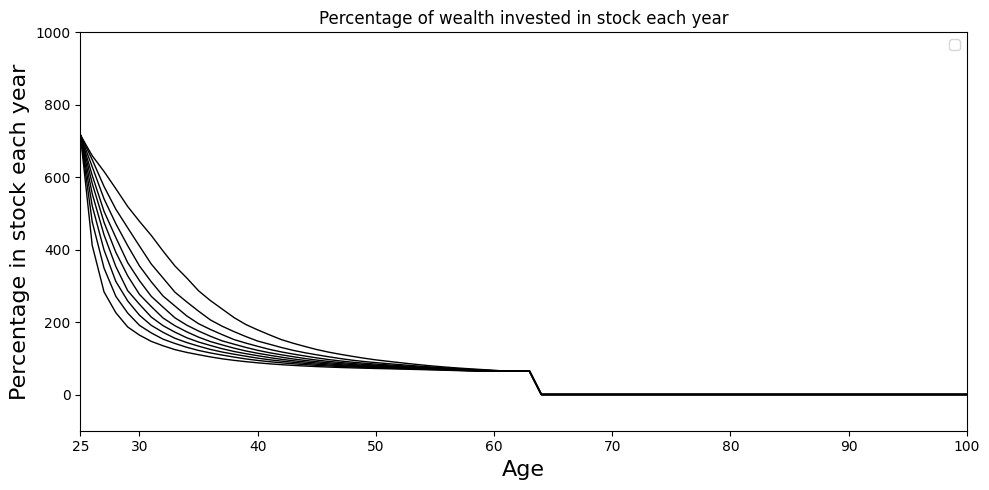}
    \end{minipage}
    \begin{minipage}{0.365\textwidth}
    \phantom{
        \includegraphics[width=1.0\textwidth]{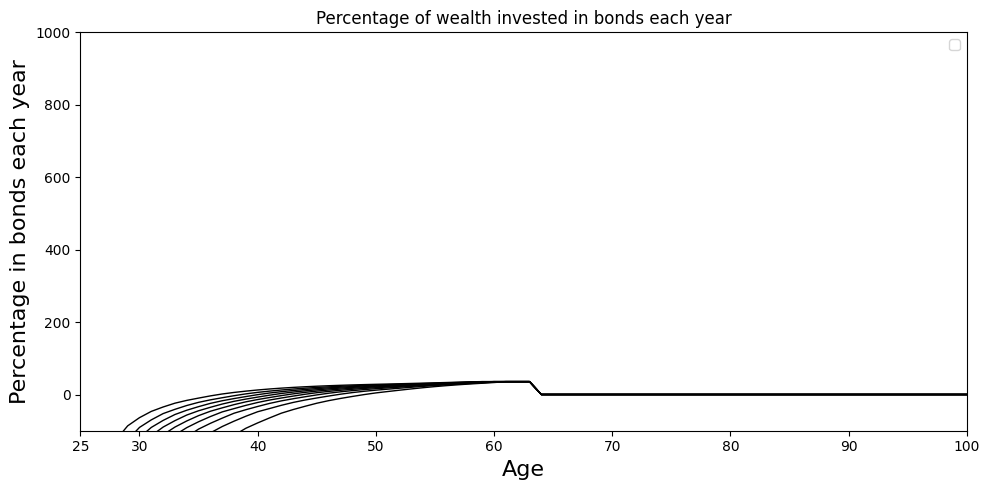}    
    }
    \end{minipage}    
        \caption{Fan diagram for the optimal DC + Annuity strategy in the Black--Scholes model, showing deciles of the percentage of wealth spent on the stock. Note that there is no investment after retirement}
    \label{fig:annuityStrategy_BlackScholes}
\end{figure}
\begin{figure}[h!tbp]
    \centering
    \begin{minipage}{0.625\textwidth}
    \centering
        \includegraphics[width=1.0\textwidth]{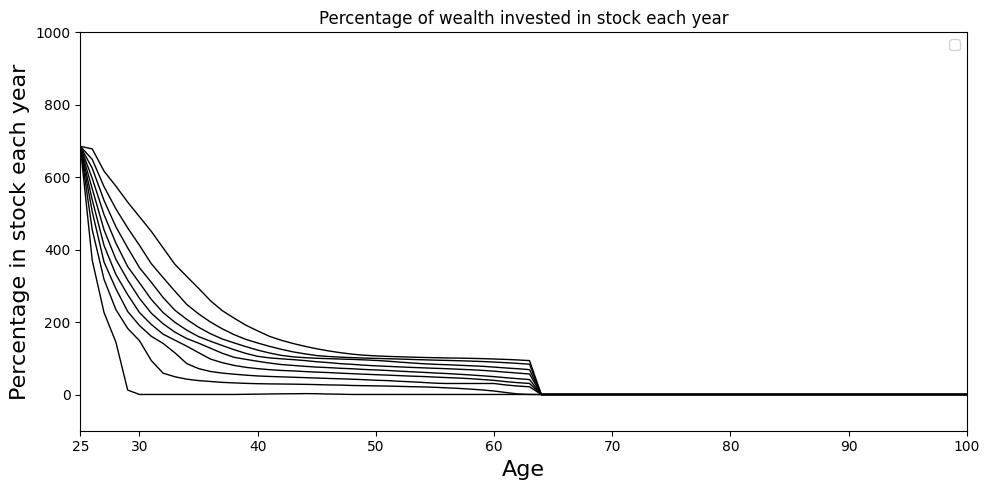}
    \end{minipage}
    \begin{minipage}{0.365\textwidth}
    \phantom{
        \includegraphics[width=1.0\textwidth]{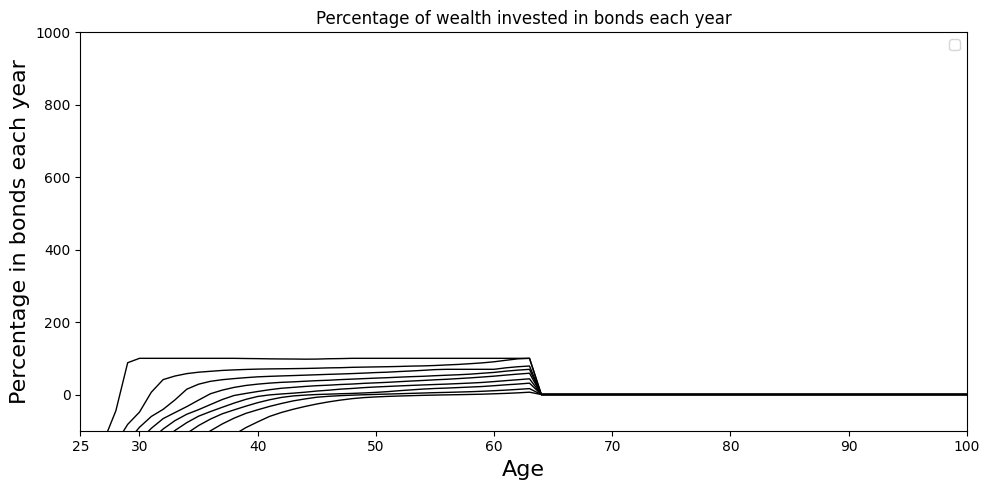}    
    }
    \end{minipage}    
        \caption{Fan diagram for the optimal DC + Annuity strategy in the SIR model, showing deciles of the percentage of wealth spent on the stock. Note that there is no investment after retirement}
    \label{fig:annuityStrategy_SIR}
\end{figure}
\begin{figure}[h!tbp]
    \centering
    \begin{minipage}{0.625\textwidth}
    \centering
        \includegraphics[width=1.0\textwidth]{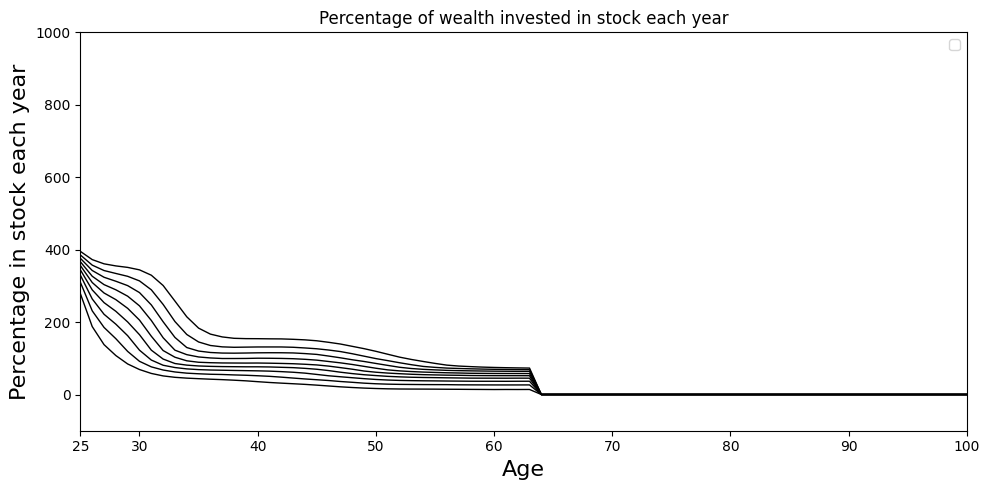}
    \end{minipage}
    \begin{minipage}{0.365\textwidth}
    \phantom{
        \includegraphics[width=1.0\textwidth]{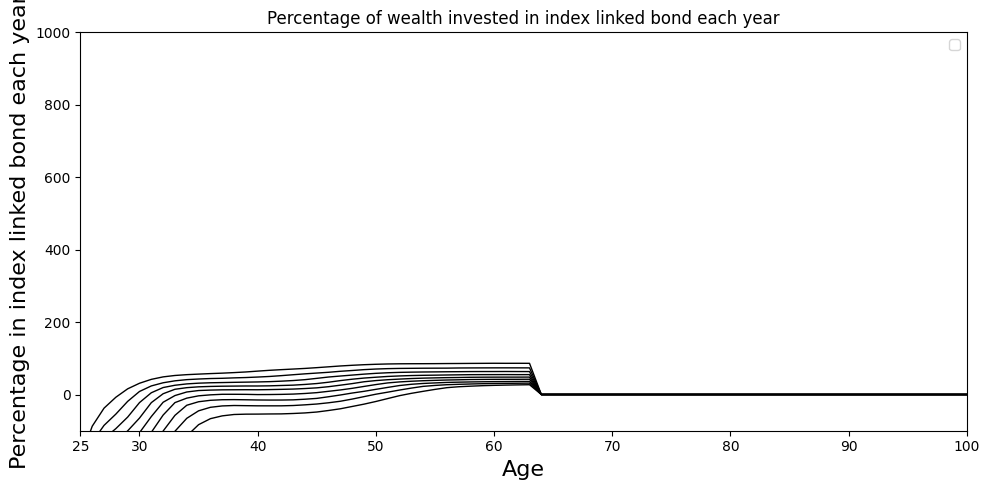}    
    }
    \end{minipage}    
        \caption{Fan diagram for the optimal DC + Annuity strategy in the richer economic model, showing deciles of the percentage of wealth spent on the stock. Note that there is no investment after retirement}
    \label{fig:annuityStrategy_ppiEsg}
\end{figure}

\pagebreak
\subsection{Flex-and-fix}

\begin{figure}[h!tbp]
    \centering
    \begin{minipage}{0.625\textwidth}
    \centering
        \includegraphics[width=1.0\textwidth]{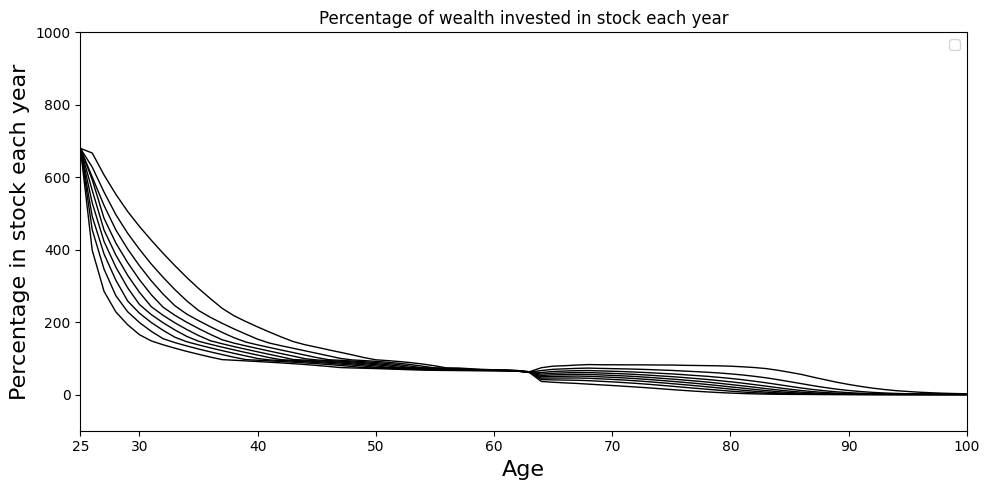}
    \end{minipage}
    \begin{minipage}{0.365\textwidth}
    \centering
        \includegraphics[width=1.0\textwidth]{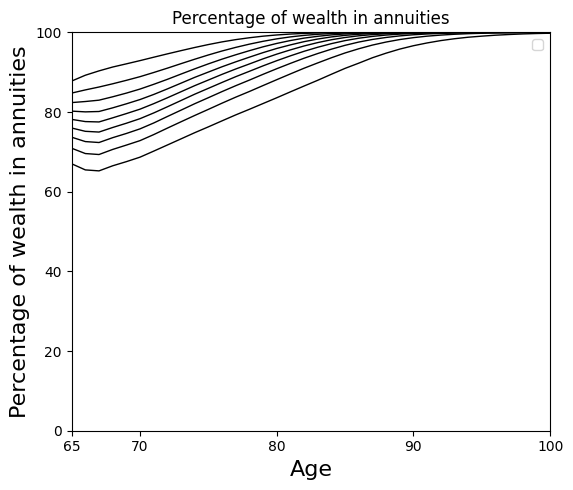}
    \end{minipage}
        \caption{Fan diagrams for the optimal flex-and-fix strategy in the Black--Scholes model, showing deciles of: the percentage of wealth spent on the stock; the percentage of wealth annuitised.}
    \label{fig:flexThenFix_BlackScholes}
\end{figure}
\begin{figure}[h!tbp]
    \centering
    \begin{minipage}{0.625\textwidth}
    \centering
        \includegraphics[width=1.0\textwidth]{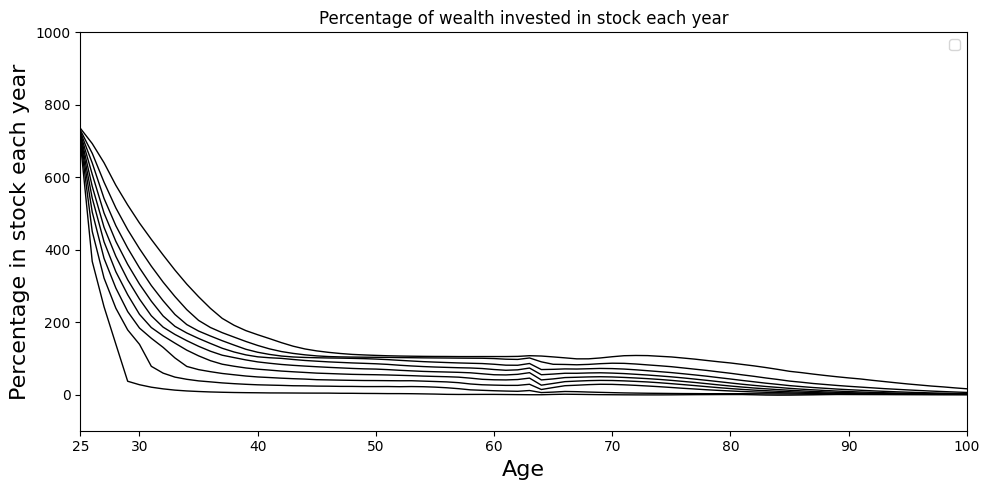}
    \end{minipage}
    \begin{minipage}{0.365\textwidth}
    \centering
        \includegraphics[width=1.0\textwidth]{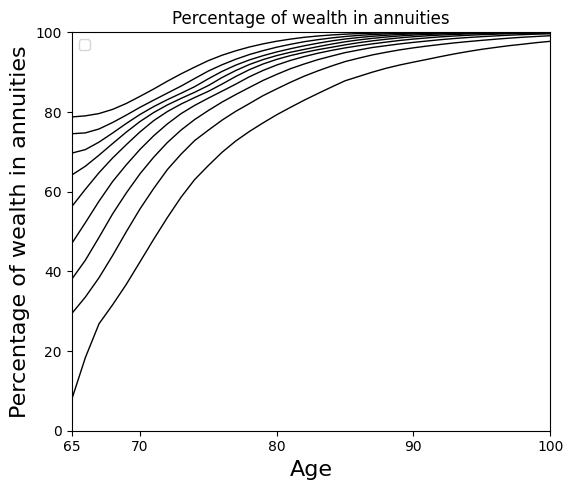}
    \end{minipage}
        \caption{Fan diagrams for the optimal flex-and-fix strategy in the SIR model, showing deciles of: the percentage of wealth spent on the stock; the percentage of wealth annuitised.}
    \label{fig:flexThenFix_SIR}
\end{figure}
\begin{figure}[h!tbp]
    \centering
    \begin{minipage}{0.625\textwidth}
    \centering
        \includegraphics[width=1.0\textwidth]{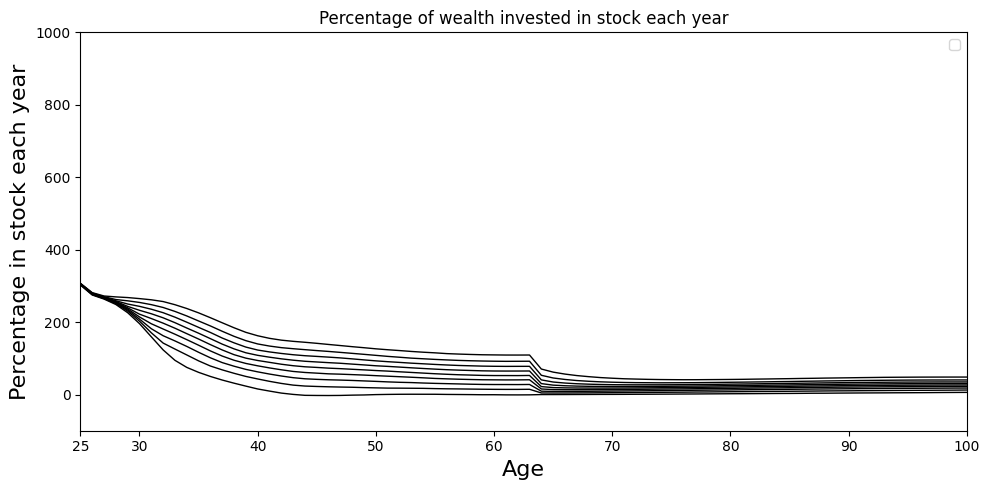}
    \end{minipage}
    \begin{minipage}{0.365\textwidth}
    \centering
        \includegraphics[width=1.0\textwidth]{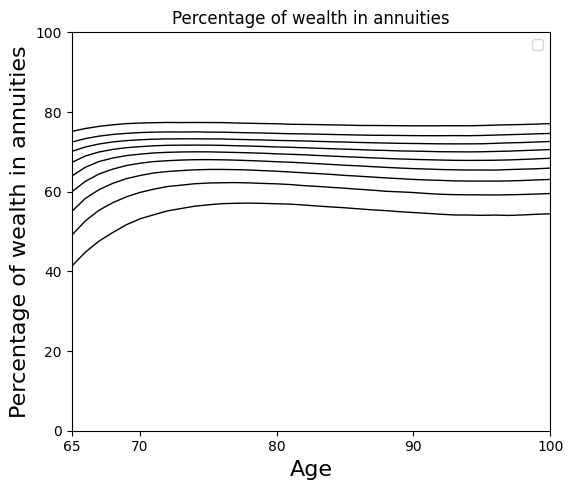}
    \end{minipage}
        \caption{Fan diagrams for the optimal flex-and-fix strategy in the richer economic model, showing deciles of: the percentage of wealth spent on the stock; the percentage of wealth annuitised.}
    \label{fig:flexThenFix_ppiEsg}
\end{figure}

\subsection{Optimal shared-indexation designs}

\begin{figure}[h!tbp]
\begin{tabular}{cc}
\includegraphics[width=0.5\textwidth]{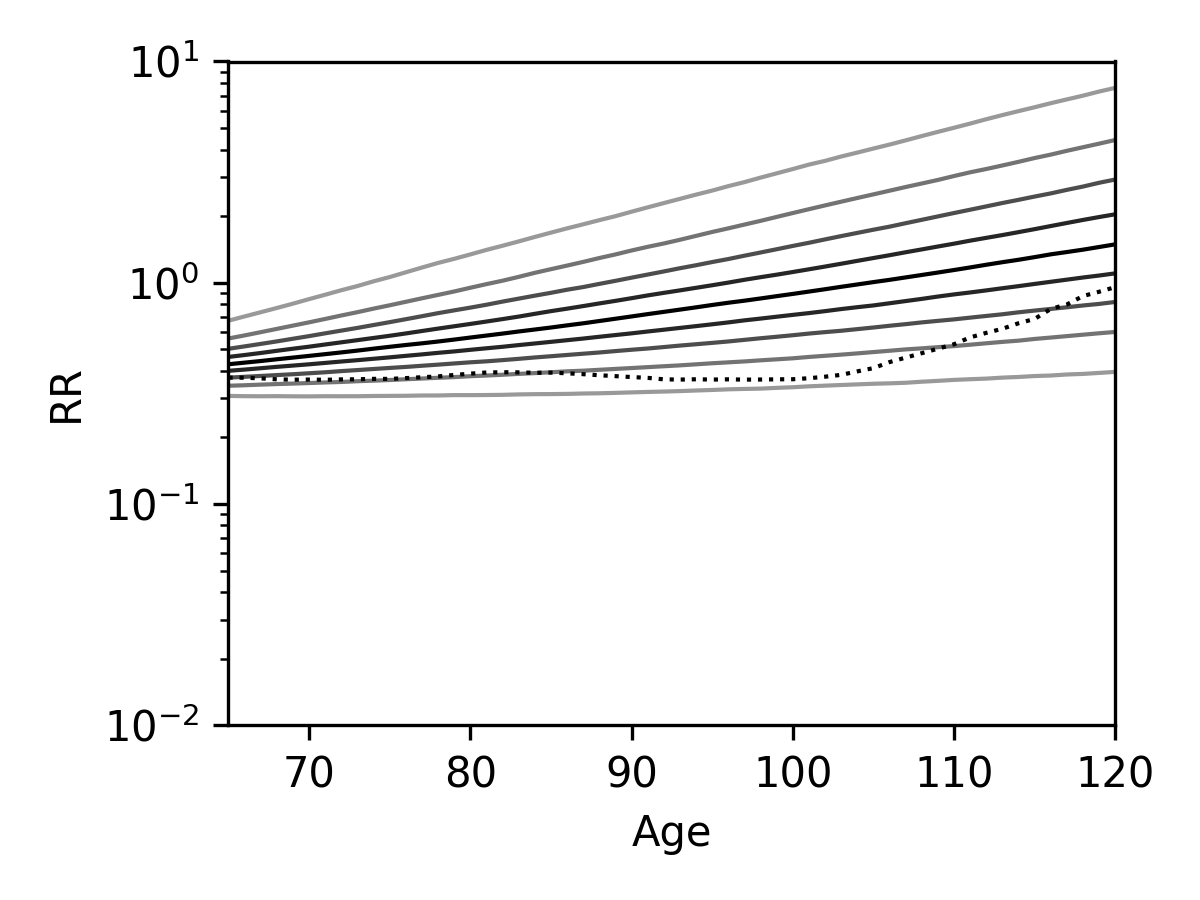} & 
\includegraphics[width=0.5\textwidth]{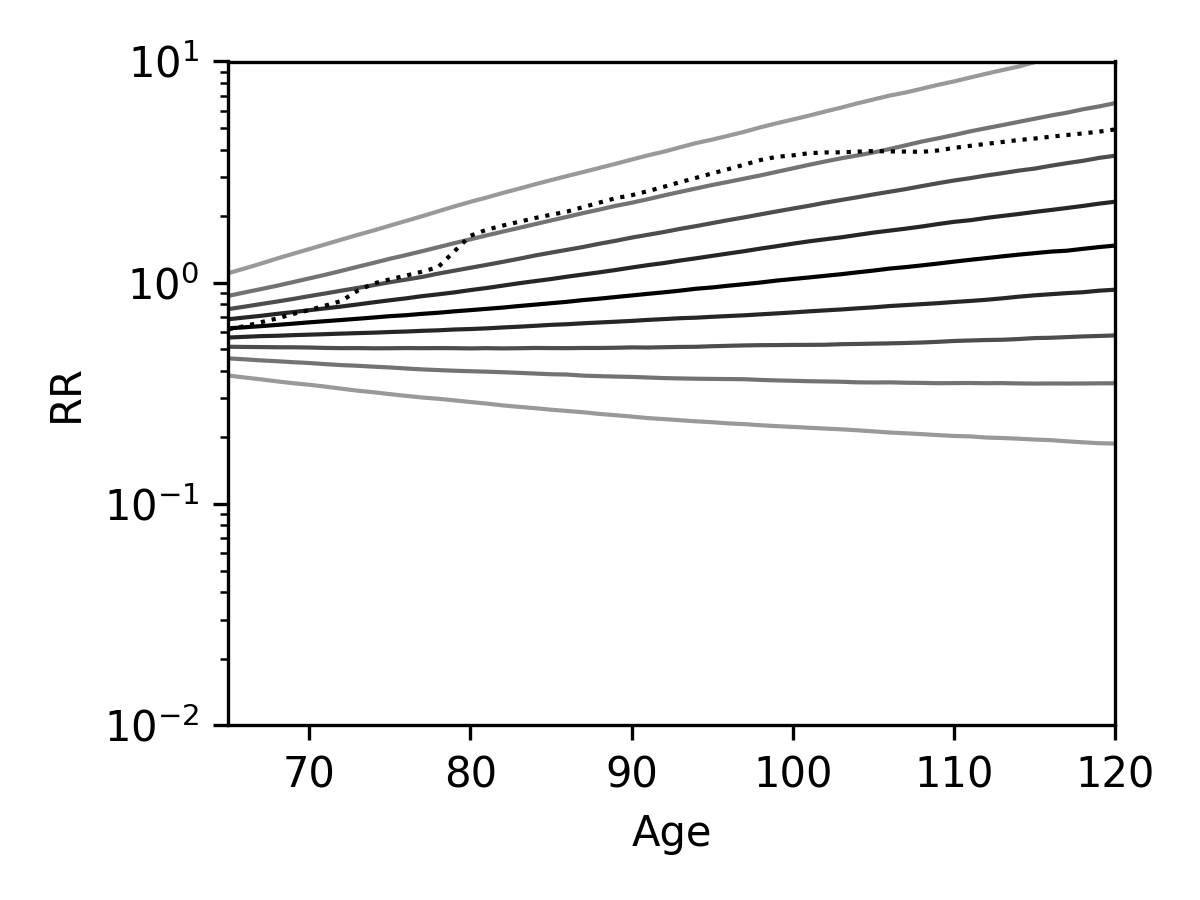} \\
(a) & (b) \\
\end{tabular}
\caption{Outcomes of the optimal shared-indexation designs in the richer economic model: (a) flat-accrual,
(b) dynamic-accrual.
Each plot
shows the deciles of the replacement ratio and
an example scenario. The generation shown in the plot
is the generation in the range (40--95) with the worst outcomes. This was generation 40 in each case.}
\label{fig:sharedIndexationFigure}
\end{figure}

\section{Effectiveness of the internal longevity market}
\label{sec:finiteFund}

In the simulations in the body of the paper we assume an infinite fund size. In this section,
we investigate how large collective drawdown schemes need to be for the longevity market to be effective. 

We simulated a sequence of collective-drawdown plans each containing 20 members. Within each plan, all members have the same age, investment and consumption strategy. We create a sequence of plans, each representing a different cohort of members. In our simulations, we treat each of these plans as though they are a separate scheme. This simulation can therefore be interpreted as both an example of how a collective-drawdown scheme might operate for a single employer with finite fund size, and as an example of how multiple small, demographically distinct schemes can effectively share a longevity market.

Some care is required to apply the strategy for an infinite fund to a finite fund. If this is done naively, the neural network can lose track of a member's wealth as the network does not receive any input providing data on how mortality has differed from central projections. To account for this, we adjust the equity risk-factors that are fed into the network to account for the noise in longevity credits which would otherwise be hidden from the network. 
Full details on how this is performed are given in section \ref{sec:appendixFiniteFundDetail}.

\begin{figure}[h!tbp]
    \begin{minipage}{0.8\textwidth}
    \centering
        \includegraphics[width=1.0\textwidth]{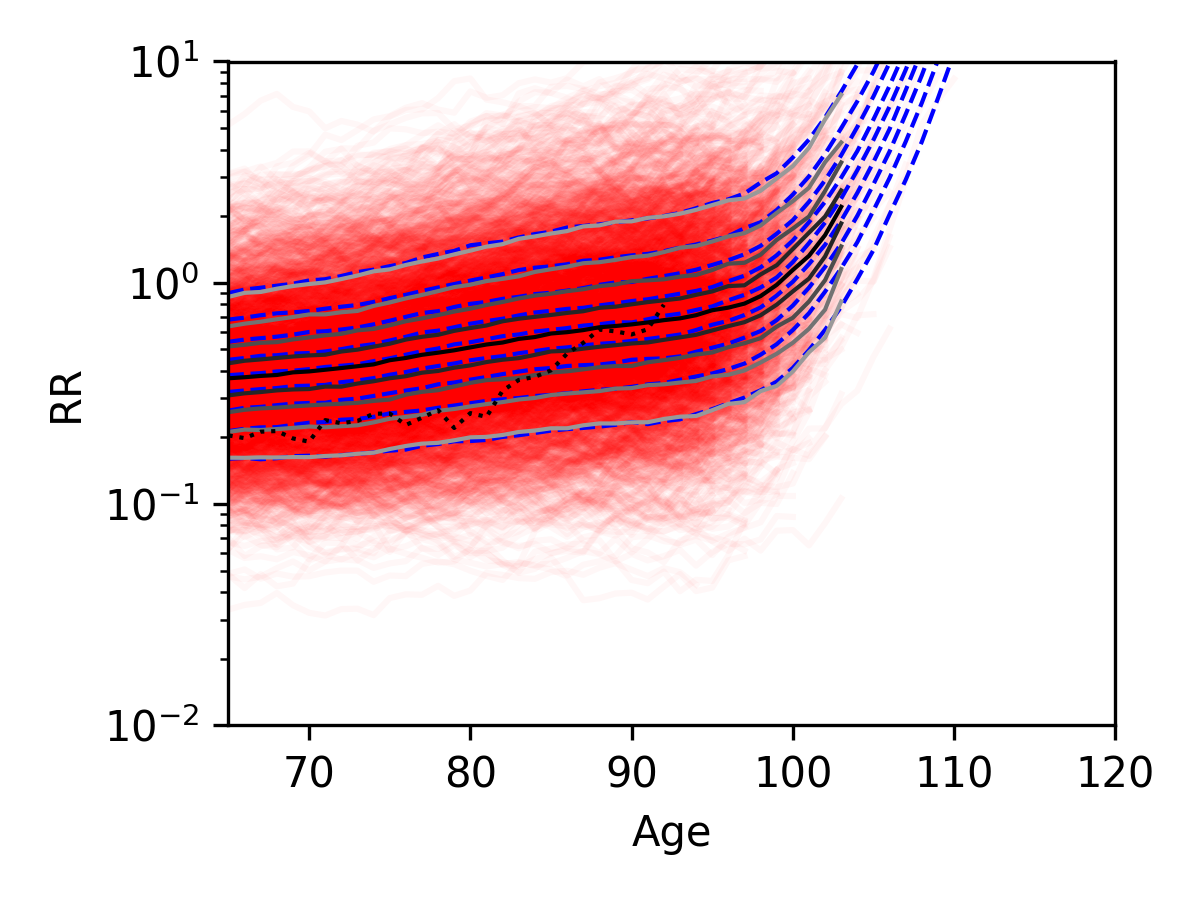}
    \end{minipage}
        \caption{Deciles of the replacement rate achieved in a collective-drawdown scheme
        with longevity pooled across generations each of size 20 (grey, solid lines).
        Individual scenarios are plotted in red. Deciles of the replacement ratio
        for an infinite fund shown in blue, dashed lines.}
    \label{fig:finiteScheme}
\end{figure}

The result for the first generation in our simulation is shown in Figure \ref{fig:finiteScheme} and is compared to the results obtained using an infinite fund. We perform 3000 simulations
to obtain these plots and only show deciles where there are at least 100 scenarios containing a survivor of the given age. These plots show the outcomes for a finite fund with 20 members in each generation. As can be seen, the plotted deciles for the finite fund are close to those of the infinite fund, and so one concludes that the tontine structure is highly effective even for small funds. The error introduced by using a finite fund compared is dwarfed by the uncertainty caused by market fluctuations.

\subsection{Using the neural network in a finite fund}
\label{sec:appendixFiniteFundDetail}

Although the network has been trained on the assumption that the fund is infinite,
in practice it will, of course, be finite. When we use the neural network to choose
investment and consumption decisions for our simulation, we modify the input representing the noise term to take account of this discrepancy. In our calculation below, $\tilde{\epsilon}_t$ will be the input used for the neural network, whereas $\epsilon_t$ will represent the actual stock increments.


The idea is to include the noise coming from
the stochastic longevity payments in the computation
of $\tilde{\epsilon}_t$, so that a neural network trained on the assumption that equation \eqref{eq:wealthInfinite} holds, will still 
correctly compute the total wealth. The true longevity payment for a finite fund is a random variable
which we denote by $P_{f,t} w_{t-}$. In a finite fund, the wealth satisfies the following modified version of equation \eqref{eq:wealthInfinite}.
\begin{equation}
    w_{t} = \eta s_t \mathbbm{1}_{t<t_{\RA}} + (1 - c_t
    \mathbbm{1}_{t\geq t_{\RA}} )(1+ P_{f,t} \mathbbm{1}_{t>t_{\RA}})w_{t-}.
\label{eq:wealthFinite}    
\end{equation}

Comparing equations \eqref{eq:wealthFinite} and \eqref{eq:wealthInfinite}, we define the modified noise process $\tilde{\epsilon}_t$ by requiring that the following equation holds
\begin{equation*}
\begin{split}
(1 + P_{f,t} \mathbbm{1}_{t>t_{\RA}}) &\exp(F(\pi_{t-\delta t}, \epsilon_{t-\delta t})) w_{t-\delta t} \\
    &=
(1 + P_{\infty,t} \mathbbm{1}_{t>t_{\RA}})  \exp(F(\pi_{t-\delta t}, \tilde{\epsilon}_{t-\delta t})) w_{t-\delta t},
\end{split}
\end{equation*}
where $F$ defines the log wealth increment.
When we pass $\tilde{\epsilon}_t$ as the noise input into the neural network which has been trained assuming an infinite fund, it
will compute the wealth of the fund at the next period
so that it matches that given by equation \eqref{eq:wealthFinite}.

\end{document}

%% file: packages.tex
\usepackage{amsthm}
\usepackage{amsmath}
\usepackage{amssymb}
\usepackage{enumerate}
\usepackage{xcolor}
\usepackage{listings}
\usepackage{multirow}
\usepackage{color}
\usepackage[pdfencoding=auto, psdextra]{hyperref}
\usepackage{doi}
\usepackage{booktabs}
\usepackage{rotating}
\usepackage{bm}
\usepackage{tablefootnote}
\usepackage[normalem]{ulem}
\usepackage{array}
\usepackage{longtable}
\usepackage{bbm}
\usepackage{cleveref}
\usepackage{tabularx}
\usepackage{multirow}
\usepackage{subcaption}
\usepackage{array}
\usepackage{makecell}
\usepackage{multirow}
\usepackage{rotating}

\newcommand{\ed}{\mathrm{d}}
\newcommand{\R}{\mathbb{R}}

\newcommand{\CPI}{\mathrm{CPI}}

\newcolumntype{L}[1]{>{\raggedright\let\newline\\\arraybackslash\hspace{0pt}}p{#1}}

\DeclareMathOperator{\argmax}

%% file: theorems.tex
\newtheorem{theorem}{Theorem}
\newtheorem{lemma}[theorem]{Lemma}

\newtheorem{proposition}[theorem]{Proposition}

\theoremstyle{definition}

\numberwithin{equation}{section}
\numberwithin{theorem}{section}